\documentclass[                     
               preprintnumbers,     
               aps,                 
               prd,                 
               superscriptaddress,      
               nofootinbib,         
               tightenlines,        
               floats,floatfix      
               ]{revtex4}

\usepackage{graphicx}
\usepackage{latexsym}
\usepackage{amsmath, amssymb}        
\usepackage{mathrsfs}
\usepackage[draft=false]{hyperref}
\usepackage{mathtools}
\usepackage[dvipsnames]{xcolor}
\usepackage{epsfig}
\usepackage{multirow}
\usepackage{tikz}
\usepackage{figsize}

\begin{document}


\title{Euler-Heisenberg Black Holes in Einsteinian Cubic Gravity}

\author{Gustavo Gutierrez-Cano}
\email{ggutierrezcano@ugto.mx}
\author{Gustavo Niz}
\email{g.niz@ugto.mx}
\affiliation{Departamento de F\'isica, Divisi\'on de Ciencias e Ingenier\'ias, Campus Le\'on, Universidad de Guanajuato, C.P. 37150, Leon, Mexico.}


\date{\today}


\begin{abstract}
We explore black hole solutions and some of its physical properties in Einstein's theory in 4D, modified by a cubic gravity term and in the presence of non-linear electrodynamics. In the context of Effective Field Theories (EFT) and under certain assumptions, these curvature and non-linear electromagnetic terms represent the first  corrections to the Einstein-Maxwell theory. We obtain static and spherically symmetric generalizations to the asymptotically flat Reissner–Nordström metric using perturbative methods, showing how an asymptotic expansion solution connects with a near horizon solution for a small coupling of the curvature correction term. Finally, we discuss how these EFT corrections change the event horizon properties and also lead to measurable effects on black hole shadows and gravitational lensing around these solutions.
\end{abstract}

\maketitle



\section{Introduction}\label{Intro}

The idea that nature splits into independent realities over distinct range scales, enables us to describe physical events on a particular scale without requiring knowledge of phenomena on other scales. For example, one may analyze the energy levels of the hydrogen atom without knowing the internal structure of the proton. In this context, the so-called Effective Field Theories (EFT) provide this framework where physical phenomena occurring at low energies are decoupled from physics at higher energies. In these EFTs an energy scale, say $\mathcal{M}$, controls the hierarchy of contributions to the theory. Moreover, the key elements to construct EFTs may be summarized as follows\cite{Pich}: i) the dynamics at long distances or low energies should not depend upon the dynamics at short distances or high energies, ii) this low energy dynamics should be described in terms of a finite set of parameters, iii) high-energy dynamics would only be included through the low-energy coupling constants and symmetries of the theory, iv) if large energy gaps exist then light scales should be set to zero while heavy scales to infinity, and finally v) the EFT should have the same infra-red behavior as the fundamental theory it comes from, if it exists.

An important example for our discussion of an EFT is the well-known Euler-Heisenberg (EH) Lagrangian\cite{HeisenbergEuler}, which results from the analysis of Quantum Electrodynamics (QED) at low energies $E_{\gamma}<<m_{e}$, where $E_{\gamma}$ and $m_{e}$ are the photon energy and electron mass, respectively. The EH theory allows for a description of the light-by-light scattering process in terms of the electromagnetic field, $A_{\mu}$. As it is usual in EFT's, the interaction terms are organized as a power series in the inverse of an energy scale, $\mathcal{M}$, which in this case corresponds to $\mathcal{M}\sim m_{e}$. In the context of QED, an effective Lagrangian that preserves gauge invariance, Lorentz symmetry, charge conjugation and parity can be written as\cite{Manohar}\footnote{Alternatively, in the literature it is possible to find\cite{Pich} $\left(F_{\alpha\beta}F^{\alpha\beta}\right)^2$ and $\left(F_{\alpha\beta}F^{\beta\rho}F_{\rho\sigma}F^{\sigma\alpha}\right)^2$ as a different tensor basis instead of $\left(F_{\alpha\beta}F^{\alpha\beta}\right)^2$ and $\left(F_{\alpha\beta}\breve{F}^{\alpha\beta}\right)^2$.}
\begin{equation}\label{LagEH}
    \mathcal{L}_{EH} = -\frac{1}{4}F_{ab}F^{ab}+\frac{g_{2}}{m_{e}^4}\left(F_{ab}F^{ab}\right)^2+\frac{\Tilde{g}_{2}}{m_{e}^4}\left(F_{ab}\breve{F}^{ab}\right)^2+\mathcal{O}\left(m_{e}^{-8}\right),
\end{equation}
where $F_{ab}=\partial_{a}A_{b}-\partial_{b}A_{a}$ and $\breve{F}^{ab}=\epsilon^{cdab}F_{cd}$ is the dual of $F^{ab}$. The low-energy coupling constants $g_{2}$ and $\Tilde{g}_{2}$ contain information of the original QED theory; whose values in the weak electromagnetic limit ($E<<1, B<<1$)
reduce to \cite{HeisenbergEuler, Euler}
\begin{equation*}
    g_{2}=\frac{\alpha^2}{90},\qquad \Tilde{g}_{2}=\frac{7\alpha^2}{90},
\end{equation*}
\noindent where $\alpha=e^{2}/(4\pi)$ is the fine structure constant and $e$ is the electron charge. The positiveness of the coupling constants is consistent with maintaining unitarity and causality in the theory \cite{Adams}. The first term of the lagrangian (\ref{LagEH}) is the usual kinetic term of Maxwell theory which contains no interactions, while the other two terms, which are the first nontrivial low-order terms in $1/m_{e}$, contain the photon self-interactions. It is important to stress that regardless of the values of the coupling constants $g_{2}$ and $\Tilde{g}_{2}$, one can determine the dependence of the effective cross section of the light-by-light scattering on the energy $E_{\gamma}$ in the limit in which the EH theory is applicable ($E_{\gamma}<<<m_{e}$). This result is a consequence of the symmetries imposed on the EFT. One may  also notice that the corrections of order $1/m_{e}^{4}$ are of opposite sign to the Maxwell Lagrangian which causes a screening of the electric charge, this will be evident later. The complete calculation of this scattering process, considering the complete QED theory, was carried out in 1951 by Karplus and Neumann\cite{KarplusNeuman}, whose low energy limit exactly recovers the result obtained using the effective theory (\ref{LagEH}).

On the other hand, general relativity can also be considered in the context of effective field theories, which formally implies the Hilbert-Einstein Lagrangian density is only a piece of a more general effective Lagrangian, which can schematically be written as
\begin{equation}\label{AGen}    \mathcal{L}=\sum_{j}\beta_{(j)}\mathcal{R}^{(j)}=\beta_{(0)}\mathcal{R}^{(0)}+\beta_{(1)}\mathcal{R}^{(1)}+\beta_{(2)}\mathcal{R}^{(2)}+\beta_{(3)}\mathcal{R}^{(3)}+\cdots,
\end{equation}
where $\mathcal{R}^{(i)}$ are polynomials of all curvature invariants of order $0$, $1$, $2$, and $3$, respectively, while the parameters $\beta_{(j)}$ are the coupling constants for each of these polynomials terms. In particular, at the lowest orders we usually consider the familiar Einstein-Hilbert Lagrangian, given by $\mathcal{R}^{(0)}=1$ and $\beta_{(0)}=M_{P}^2\Lambda$ (with $\Lambda$ the cosmological constant and $M_{P}$ the Planck mass) at zeroth order, and $\mathcal{R}^{(1)}=R$ (the Ricci scalar) and $\beta_{(1)}=M_{P}^2/2$ at first order. The construction of higher order gravity theories in four dimensions is not trivial, because the field equations often contain derivatives higher, in order, than two for the metric field, $g_{\mu\nu}$, which usually exhibit Ostrogradski instabilities\cite{Ostro}. To circumvent this problem Lanczos\cite{Lanczos} and Lovelock\cite{Lovelock} constructed gravitational theories whose equations of motion remain second order in derivatives of $g_{\mu\nu}$. Lovelock showed that, in four dimensions, the only gravitational theory (up to boundary terms) that allows second order field equations is Einstein's theory with a cosmological constant, $\mathcal\beta_{(0)}{R}^{(0)}+\beta_{(1)}\mathcal{R}^{(1)}=M_{P}^{2}/2\left(\Lambda+R\right)$, while for dimension $D\geq 5$ there is an additional quadratic order term of curvature, $\mathcal{R}^{(2)}=\chi_4$, known as the Gauss-Bonnet term which is also dynamic. For $D\geq 7$, a new non-topological Lovelock cubic curvature term shows up, given by $\mathcal{R}^{(3)}=\chi_{6}$. These two terms $\chi_{4}$ and $\chi_{6}$ in four dimensions correspond to Euler densities and do not contribute to the dynamics of the field equations, unless they are coupled to other matter fields such as a scalar field or a vector field (see for example\cite{Metsaev, Gasperini, Antoniou, Aoki}).

Recently, a different cubic theory of gravity in four dimensions has been proposed, which contributes to the dynamics of the field equations without a matter coupling. This construction is base on Lovelock's quadratic densities and the following desired properties i) General Relativity is recovered by considering the coupling constants $\beta^{(2)}=0=\beta^{(3)}$, ii) on maximally symmetric backgrounds, the linearized theory (\ref{AGen}) propagates the same degrees of freedom as linearized Einstein gravity, and iii) the field equations derived from (\ref{AGen}) admit Schwarzschild-like solutions, which are characterized by a single metric function\footnote{By Schwarzschild-like solutions we mean static spherically symmetric metrics that satisfy the condition $g_{tt}g_{rr}=-1$ while the rest of the components correspond to the line element on a surface of constant scalar curvature and we focus on spherical topology, see eq. (\ref{metric}).}. To quadratic order, the most general polynomial $\mathcal{R}^{(2)}$ satisfying these conditions is the Gauss-Bonnet term, $\chi_{4}$, which is not dynamic in 4D; while to cubic order the polynomial $\mathcal{R}^{(3)}$ is written as follows\cite{Kubiznak}
\begin{equation}
    \mathcal{R}^{(3)} = \frac{1}{4}\chi_{6}+\Pi-8\mathcal{C},
\end{equation}
where $\chi_{6}$ is the six-dimensional Euler density corresponding to the cubic Lovelock term, while $\Pi$ and $\mathcal{C}$ are two new cubic invariant polynomials. Recall that for $D=4$ the Lovelock terms $\chi_{4}$ and $\chi_{6}$ vanish and so does the term $\mathcal{C}$, therefore the lagrangian (\ref{AGen}) in four dimensions up to cubic terms of curvature, known as Einsteinian Cubic Gravity (ECG), which satisfies the previously listed properties is written as\cite{BuenoCano1}.
\begin{equation}{\label{AECGF}}
    \mathcal{L}=\frac{M_{P}^{2}}{2}\left(R+\beta\Pi\right).
\end{equation}

As mentioned above, effective field theories describe the dynamics of low energies only up to a certain energy scale $\mathcal{M}$. In the case of gravitational effective theories, when they exceed this energy scale, they exhibit Ostrogradski instabilities. For the case of ECG theory, these unstable modes have been found while considering perturbations over static and spherically symmetric solutions \cite{DeFelice}, though these modes have an associated energy scale above the regime of validity of the EFT \cite{JimenezCano}. In detail, the cubic terms ($\beta\Pi$) of the gravitational Lagrangian (\ref{AECGF}) are suppressed with respect to the Einstein-Hilbert term by $1/\mathcal{M}^2$ while the coupling constant $\beta$ is related to the cutoff scale as $\mathcal{M}=1/\beta^{1/4}$. The existence condition for these instabilities is $r^{3}<<\beta r_{h}\mathcal{M}^2$ \cite{DeFelice}, which in the case of $\mathcal{M}=1/\beta^{1/4}$ reduces to $r^{3}<<\beta^{1/2}r_{h}$. Therefore, this condition indicates that these cubic correction terms in (\ref{AECGF}) have to be smaller than the linear term outside the event horizon of a black hole in order to avoid pathologies\footnote{Note that our notation is equivalent to that in \cite{DeFelice} when $M=\mathcal{M}$ and $\alpha=\beta$.}. In the present work, we consider our solutions are valid in the regime where $|\beta\Pi|<<|R|$, i.e. in the regime where the EFT is valid, hence avoiding instabilities. This approach would exclude all non-perturbative effects that can generate corrections to the Einstein-Hilbert action, as well as quantum effects that might solve the presence of ghosts \cite{Hawking} \footnote{For an alternative approach to study quantum corrections, applied to the Schwarzschild metric, under a gravitational EFT see \cite{Battista}}.

From a different perspective, the data obtained by the Event Horizon Telescope (EHT) on the observations of the supermassive black holes M87*\cite{Akiyama1,Akiyama2} in the galaxy M87 and of the black hole SgrA*\cite{Akiyama3,Akiyama4} located in the center of the Milky Way, together with the detection of gravitational waves by LIGO and VIRGO\cite{LIGO}, open the door to test the different properties of compact objects, whether in General Relativity or in alternative theories of gravity. 
In this context, one may explore physical signals associated to black holes in the ECG and EH theories, namely the Lagrangian (\ref{LagEH}) and (\ref{AECGF}), and how these properties are sensitive to the curvature and electromagnetic field higher order corrections.

The article is organized as follows. In section \ref{AEQ} we present the action and field equations of the effective field theories of Einsteinian Cubic Gravity and Euler-Heisenberg-like nonlinear electrodynamics, while in the following Section \ref{Solution} construct  black hole solutions as series expansions. In section \ref{Properties}, we study the motion of massive and massless test particles around the black hole solutions and subsequently construct the shadow contour. We end up that section discussing the modification to the lensing signal around these black holes. Finally, the section \ref{Fcomm} is reserved for conclusions. Notice that throughout this work we use the natural units of $c=1$ and $G=1$. 


\section{Action and Equations}\label{AEQ}

In four dimensions, a non-topological cubic gravity theory, that admits Schwarzschild-like black hole solutions and maintains the same degrees of freedom as General Relativity in a maximally symmetric spacetime coupled to non-linear electrodynamics, can be casted in terms of the following the action
\begin{equation}\label{Action}
S=\int{d^4x\sqrt{-g}\left[\frac{1}{16\pi}\left(R+\beta\Pi\right)-\frac{1}{4\pi}\left(\frac{1}{2}P^{ab}F_{ab}-\mathscr{H(P)}\right)\right]},
\end{equation}
 where $R$ is the Ricci scalar, $\beta$ is the cubic term coupling constant, while $\Pi$ is the cubic correction given by 
\begin{equation}\label{InvP}
\Pi=12{{{R_{a}}^{c}}_{b}}^{d}{{{R_{c}}^{e}}_{d}}^{f}{{{R_{e}}^{a}}_{f}}^{b}+{R_{ab}}^{cd}{R_{cd}}^{ef}{R_{ef}}^{ab}-12R_{abcd}R^{ac}R^{bd}+8{R_{a}}^{b}{R_{b}}^{c}{R_{c}}^{a}.
\end{equation}
The last two term in (\ref{Action}) determine the nonlinear behavior of the electromagnetic field $A_{a}$, in which a simple way to include nonlinear effects in electrodynamics is to consider a Lagrangian that is built out as a general function of the quadratic invariant $\mathscr{F}=\frac{1}{4}F_{ab}F^{ab}$. However, there is a different formulation, developed by Plebanski\cite{Plebanski,SalazarGarciaPlebanski} ,that introduces a structural component $\mathscr{H(P)}$, that is a function of the invariants formed with the antisymmetric tensor $P^{ab}$, namely
\begin{equation}
\mathscr{P}=\frac{1}{4}P_{ab}P^{ab},    
\end{equation}  
which in turn is related to the Lagrangian by a Legrende transformation\footnote{In general the structure function $\mathscr{H}$ can depend on the invariants $\mathscr{P}$ and $\mathscr{Q}$, where this second invariant is defined as $\mathscr{Q}=\frac{1}{4}P_{ab}\breve{P}^{ab}$, with $\breve{P}^{ab}=\epsilon^{cdab}P_{ab}$ being the dual of $P^{ab}$.}. This formulation, also known as the $\mathscr{HP}$-framework of electrodynamics, has proven to be suitable for finding purely electrical solutions\cite{MagosBreton,AmaroMacias}, while the usual $\mathscr{LF}$-framework formulation tends to be more suitable for finding purely magnetic solutions\cite{Yajima}. The advantage of using the $\mathscr{HP}$-framework formalism is that the variation of the action (\ref{Action}) with respect to the electromagnetic field $A_{a}$ gives a Maxwell-like equation 
\begin{eqnarray}\label{MaxEQ}
\nabla_{a} P^{ab}=0,
\end{eqnarray}
which are linear in $P^{ab}$. The nonlinearity of the electromagnetic theory is hidden in the \textit{constitutive relations}, which are obtained from the variation of the action (\ref{Action}) with respect to the antisymmetric tensor $P^{ab}$, namely
\begin{equation}\label{ConstEQ}
F_{ab}=\mathscr{H_P}P_{ab},
\end{equation}
with $\mathscr{H_P}=\frac{\partial\mathscr{H}}{\partial \mathscr{P}}$. Linear Maxwell's electrodynamics is recovered using the choice $\mathscr{H(P)}=\mathscr{P}$. It is important to stress that both formalisms are not equivalent in many cases, as it happens in the standard Legendre transformations. The equivalence depends on the invertibility of the conjugate variables.

After stating the departure theory, the next step is to derive the field equations in the usual way. A variation of (\ref{Action}) with respect to the metric yields
\begin{equation}\label{EinsEQ}
    \mathcal{E}_{ab}:=P_{acde}R_{b}^{\ cde}-\frac{1}{2}g_{ab}\mathcal{L}_{ECG}-2\nabla^{c}\nabla^{d}P_{acdb}-8\pi T_{ab}=0,
\end{equation}
where $\mathcal{L}_{ECG}$ is the Lagrangian of Einsteinian Cubic Gravity, and $P_{abcd}\equiv\frac{\partial\mathcal{L}_{ECG}}{\partial R^{abcd}}$ is given
\begin{eqnarray}\label{PabcdTen}
P_{abcd}&&=g_{a[c}g_{b]d}+6\beta\Big[R_{ad}R_{bc}-R_{ac}R_{bd}+g_{bd}R_{a}^{\ e}R_{ce}-g_{ad}R_{b}^{\ e}R_{ce}-g_{bc}R_{a}^{\ e}R_{de}+g_{ac}R_{b}^{\ e}R_{de}-g_{bd}R^{ef}R_{aecf}\nonumber\\
&& \left.+g_{bc}R^{ef}R_{aedf}+g_{ad}R^{ef}R_{becf}-3{{{R_{a}}^{e}}_{d}}^{f}R_{becf}-g_{ac}R^{ef}R_{bedf}+3{{{R_{a}}^{e}}_{c}}^{f}R_{bedf}+\frac{1}{2}R_{ab}^{\ \ ef}R_{cdef}\right].
\end{eqnarray}
Finally, the energy-momentum tensor is written as 
\begin{equation}\label{EMTen}
4\pi T_{ab}={\mathscr{H_P}}P_{ac}P_{b}^{\ c}-g_{ab}\left(2\mathscr{P}{\mathscr{H_P}}-\mathscr{H}\right).
\end{equation}
As previously explained, in this work we focus on the so-called electric configurations, in which only the usual electric components of the field strength are turn on. The counterpart to these solutions, the magnetic configurations, can be obtained via the simple electromagnetic duality of nonlinear electrodynamics (see for example \cite{GibbonsRasheed}). For the electric case, the structure function $\mathscr{H}$ may be written as 
\begin{equation}\label{MNLE}
\mathscr{H(P)}=g_{1}\mathscr{P}+\frac{g_{2}}{2}\mathscr{P}^2,
\end{equation}
where we add a new coupling constant $g_{1}$ which can only take values $\{0, 1\}$, whose discrete values allow for a study of the solutions without the linear contribution. In contrast, the coupling constant $g_{2}$ (known as the Euler-Heisenberg parameter) is required to be positive, $g_{2}\geq 0$, in order to avoid violations of causality and unitarity \cite{Adams}. It is straightforward to recover Maxwell's theory for $g_{1}=1$ and $g_{2}=0$.

Before describing our search for solutions to the field equations (\ref{MaxEQ})-(\ref{EinsEQ}), it is interesting to review some of the results found in the literature about black holes found within these two EFTs, which have been mostly studied separately. For example, the search of black hole solutions in General Relativity (GR) coupled to the Euler-Heisenberg electrodynamics were found by Yajima and Tamaki  for magnetic charges only \cite{Yajima}, while the equivalent electric solutions were discovered in \cite{MagosBreton,AmaroMacias}, where the authors adopted the Plebanski $\mathscr{H(P)}$ formulation. Efforts to include rotation were also pursued, as in \cite{NoraRot}. Moreover, different properties of some of these solutions have been analyzed, such as the geodesic structure\cite{AmaroMacias}, thermodynamic stability\cite{MagosBreton} and the shadow contour\cite{Allahyari}. Even beyond the usual GR structure, EH electrodynamics has also been studied in the context of Modified Gravity\cite{Stefanov,GuerreroRubiera}. On the other hand, black holes with an ECG correction term which generalised the Schwarzchild and Reissner-Norsdtröm metrics were described in\cite{BuenoCano2,HennigarMann} as well as approximate analytical solutions using fast-converging continued-fraction expansions \cite{Sajadi}. Moreover, the authors in \cite{AdairBuenoCano} and \cite{CanoPeriniguez}, respectively, obtained slowly and extremal rotating black holes with these cubic gravitational models. Finally, there are other interesting physical applications of these cubic curvature corrections beyond the common black hole solutions, such as in the inflationary epoch the universe\cite{Arciniega1,Arciniega2,Arciniega3}, holographic studies \cite{BuenoCanoRuiperez,EdelsteinGrandi}, observational deviations from GR\cite{Poshteh1,Poshteh2}, and other compact object such as NUT \cite{NUTs} and wormholes\cite{Mehdizadeh}  solutions.

After this small review of the literature, it is time to present the static, spherically symmetric solution to the field equations (\ref{MaxEQ})-(\ref{EinsEQ}), using the structural function (\ref{MNLE}) which present electric configurations.


\section{Solving Field Equations}\label{Solution}

Given that the ECG theory admits Schwarzschild-like solutions in four dimensions, without loss of generality, we start the solution's search by assuming a static, spherically symmetric metric ansatz, characterized by a single metric function $K(r)$, namely\footnote{Static spherically symmetric spacetimes in general can have a lapse function and a term proportional to $dtdr$. The latter can be eliminated by a new time coordinate $T(t,r)=t+\psi(r)$, while the field equations (\ref{EinsEQ}) set the lapse function to be a constant, which then can be absorbed in a redefinition of the $T$-coordinate, see \cite{BuenoCano2}.}
\begin{equation}\label{metric}
ds^2=-K(r)dt^2+\frac{dr^2}{K(r)}+r^2d\theta^2+r^2\sin^2{(\theta)}d\phi^2.
\end{equation} 
The only non-vanishing components of the antisymmetric $P_{ab}$ tensor compatible with this metric ansatz are $P_{tr}$ and $P_{\theta\phi}$\cite{GarciaGut}. As previously mentioned, we are only considering electrically charged configurations, thus the $P_{ab}$ tensor should have an antisymmetric form\footnote{Square brackets indicate antisymmetrization of the $a$ and $b$ indices, i.e. $A_{[ab]}=\frac{1}{2}(A_{ab}-A_{ba})$.}  $P_{ab}=2\delta^{t}_{[a}\delta^{r}_{b]}D(r)$, with $D(r)$ an arbitrary function. Under this condition, equation (7) to be easily integrated, resulting in
\begin{equation}\label{TPmn}
P_{ab}=2\delta^{t}_{[a}\delta^{r}_{b]}\frac{q}{r^2}\rightarrow\mathscr{P}=-\frac{q^2}{2r^4},
\end{equation}
where the only integration constant, $q$, is the electric charge. The associated electric field could be easily obteined from the constitutive relations (\ref{ConstEQ}), $E(r) = F_{tr} = \mathscr{H_P}P_{tr}$, resulting in
\begin{equation}\label{EField}
F_{ab}=2\delta^{t}_{[a}\delta^{r}_{b]}\left(1-\frac{g_{2}}{g_{1}}\frac{q^2}{2r^4}\right)\frac{g_{1}q}{r^2}.
\end{equation}
One should notice that the electric field at a fixed radius is smaller than the one obtained on the equivalent solution using only the linear Maxwell term and this is due to the effect of the screening charge from the vacuum polarization. Also, the divergence at $r=0$ remains present, but now it is stronger since the nonlinear contribution diverges as $\mathcal{O}(r^{-6})$, and has the opposite sign to that of the Maxwell field. At the level of the solution, it is straightforward to recover the Maxwell result by taking $g_{2}\rightarrow 0$.

A master equation which already contains the electric field non-vanishing components is obtained by substituting the metric ansatz (\ref{metric}), the function $\mathscr{H(P)}$ from expression (\ref{MNLE}) and the tensor $P_{ab}$ in equation (\ref{TPmn}) into the purely radial component of the Einstein equations (\ref{EinsEQ}), which after some manipulation reduces to
\begin{eqnarray}\label{EQrr}
\mathcal{E}_{r}^{\ r}&&:=\frac{rK'+K-1}{r^2}+\beta\left[\frac{6KK'''}{r^3}\left(rK'-2K+2\right)+\frac{6K{K''}^2}{r^2}+\frac{24KK''}{r^4}\left(K-rK'-1\right)\right.\nonumber\\
&& \left.+\frac{6K'^2}{r^4}\left(4K-1\right)-\frac{24KK'}{r^5}\left(K-1\right)\right]+\frac{g_{1}q^2}{r^4}-\frac{g_{2}q^4}{4r^8}=0.
\end{eqnarray}
The other components of the Einstein equation (\ref{EinsEQ}) are either zero, or equivalent to this last expression or its derivatives. In the following,  we search for solutions of this master equation using perturbative methods.

\subsection{Asymptotic Solution}\label{ASSec}

We start by analyzing the behavior of the solution at large $r$. For this purpose we propose a series expansion in inverse power of $r$ of the metric function $K(r)$, namely
\begin{equation}
K_{\infty}(r)=1+\sum_{n=1}^{\infty}{\frac{c_{n}}{r^{n}}}.
\end{equation}
Inserting this expression into the master equation (\ref{EQrr}), and demanding that the solution should be satisfied order by order, the following coefficients are obtained
\begin{equation*}
\underbrace{c_{2}=g_{1}q^2,\quad c_{6}=-\frac{g_{2}q^4}{20}}_{\mathcal{O}(\beta^0)}\underbrace{+54c_{1}^2\beta,\quad\cdots,\quad c_{11}=-\frac{252}{5}c_{1}g_{2}q^4\beta}_{\mathcal{O}(\beta)}\underbrace{+\frac{163296}{3}c_{1}^3\beta^2+,\quad\cdots}_{\mathcal{O}(\beta^2)},
\end{equation*}
where the under-brackets group the terms at each order in the coupling constant of the cubic gravitaional interaction $\beta$. We observe that the coefficient $c_{1}$ is a free parameter, which in the limit of General Relativity (i.e. $\beta\rightarrow 0$) is associated to the Schwarszhild mass via $c_{1}=-2M$. Considering higher orders of $c_{n}$, the asymptotic solution can be casted as
\begin{eqnarray}\label{SolAsym}
K_{\infty}(r)&&=1-\frac{2M}{r}+g_{1}\frac{q^2}{r^2}-g_{2}\frac{q^4}{20r^6}+\beta\left[-\frac{108M}{r^5}\left(-\frac{2M}{r}+g_{1}q^2\frac{2}{r^2}-g_{2}q^4\frac{7}{15r^6}\right)\right.\nonumber\\
&& +g_{1}q^2\frac{108}{r^6}\left(-\frac{2M}{r}+g_{1}q^2\frac{16}{9r^2}-g_{2}q^4\frac{16}{45r^6}\right)-g_{2}q^4\frac{126}{5r^{10}}\left(-\frac{2M}{r}+g_{1}q^2\frac{32}{21r^2}-g_{2}q^4\frac{24}{105r^6}\right)\nonumber\\
&& +M^2\frac{184}{r^6}\left(-\frac{2M}{r}+g_{1}q^2\frac{57}{23r^2}-g_{2}q^4\frac{237}{230r^6}\right)-g_{1}q^2M\frac{228}{r^7}\left(-\frac{2M}{r}+g_{1}q^2\frac{28}{19r^2}-g_{2}q^4\frac{99}{95r^6}\right)\nonumber\\
&& \left. +\left(g_{1}q^2\right)^2\frac{168}{r^8}\left(-\frac{2M}{r}+g_{1}q^2\frac{19}{21r^2}-g_{2}q^4\frac{59}{140r^6}\right)+\left(g_{2}q^4\right)^2\frac{351}{50r^{16}}\left(-\frac{2M}{r}+g_{1}q^2\frac{127}{117r^2}-g_{2}q^4\frac{23}{780r^6}\right)\right]\nonumber\\
&& +\beta^2\left[M^2\frac{217728}{r^{10}}\left(-\frac{2M}{r}+g_{1}q^2\frac{10}{3r^2}-\mathcal{O}(r^{-6})\right)-g_{1}q^2M\frac{362880}{r^{11}}\left(-\frac{2M}{r}+g_{1} q^2\frac{141}{35r^2}-\mathcal{O}(r^{-6})\right)\right.\nonumber\\
&& \left.+M^2\frac{896448}{r^{12}}\left(-\frac{2M}{r}+\mathcal{O}(r^{-2})\right)-M^3\frac{886464}{r^{11}}\left(-\frac{2M}{r}+g_{1}q^2\frac{1312}{171r^2}-\mathcal{O}(r^{-6})\right)+\mathcal{O}(r^{14})\right]+\cdots\nonumber\\
\end{eqnarray}
We note that the first four terms correspond to the GR solution\cite{MagosBreton,AmaroMacias}, while the first corrections due to the presence of the cubic curvature terms appear to order $\mathcal{O}(r^{-6})$; as well as the first contribution from the nonlinear electrodynamics. One should also notice that the curvature corrections can be written as a powers series on $\beta$ with alternating signs. Moreover, the contributions proportional to the mass $M$ keep an order $\mathcal{O}(r^{-1})$ with respect to the linear electrodynamics term $g_{1}q^2$, and the latter has an additional $\mathcal{O}(r^{-4})$ factor when compared to to the nonlinear electrodynamics contribution $g_{2}q^{4}$.

\subsection{Weak Coupling Limit}\label{WCSec}

Since the solution (\ref{SolAsym}) can be written as a power series in $\beta$, it is possible to consider the cubic curvature term as a small perturbation of General Relativity, which would be consistent with the treatment of ECG theory as a gravitational EFT. The metric function $K(r)$ written in terms of the coupling constant $\beta$, which is considered to be small, can expanded as the following series
\begin{equation}\label{beta.expansion}
K_{\beta<<1}(r)=\sum_{n=0}^{\infty}{\beta^{n}k_{n}(r)}.
\end{equation}
If this series is substituted into the $\mathcal{E}_{r}^{\ r}$ component of the ECG equations (\ref{EQrr}), it yields a first order differential equation for each $k_{n}(r)$ at each order in $\beta$. These equations are easily integrated to arbitrary order in $\beta$, resulting in the following expressions for the first two order 
\begin{eqnarray}\label{SolWAco}
k_{0}(r)&&=1-\frac{2M}{r}+g_{1}\frac{q^2}{r^2}-g_{2}\frac{ q^4}{20r^6},\label{SolWAco0}\\
k_{1}(r)&&=-\frac{108M}{r^5}\left(-\frac{2M}{r}+g_{1}q^2\frac{2}{r^2}-g_{2}q^4\frac{7}{15r^6}\right)+g_{1}q^2\frac{108}{r^6}\left(-\frac{2M}{r}+g_{1}q^2\frac{16}{9r^2}-g_{2}q^4\frac{16}{45r^6}\right)\nonumber\\
&& -g_{2}q^4\frac{126}{5r^{10}}\left(-\frac{2M}{r}+g_{1}q^2\frac{32}{21r^2}-g_{2}q^4\frac{24}{105r^6}\right)+M^2\frac{184}{r^6}\left(-\frac{2M}{r}+g_{1}q^2\frac{57}{23r^2}-g_{2}q^4\frac{237}{230r^6}\right)\nonumber\\
&& -g_{1}q^2M\frac{228}{r^7}\left(-\frac{2M}{r}+g_{1}q^2\frac{28}{19r^2}-g_{2}q^4\frac{99}{95r^6}\right)+\left(g_{1}q^2\right)^2\frac{168}{r^8}\left(-\frac{2M}{r}+g_{1}q^2\frac{19}{21r^2}-g_{2}q^4\frac{59}{140r^6}\right)\nonumber\\
&& +\left(g_{2}q^4\right)^2\frac{351}{50r^{16}}\left(-\frac{2M}{r}+g_{1}q^2\frac{127}{117r^2}-g_{2}q^4\frac{23}{780r^6}\right),\label{SolWAco1}
\end{eqnarray}
(while the solution for $k_{2}(r)$ is shown in Appendix \ref{SmCoup}).  From the previous expressions, we can see that the linear term $k_{0}(r)$ corresponds to the solution of General Relativity\cite{MagosBreton,AmaroMacias}, while the first correction $k_{1}(r)$ is precisely the $\mathcal{O}(\beta)$ correction that we had already found in the asymptotic solution of equation (\ref{SolAsym}). The fact that the same expression is obtained for the asymptotic case (\ref{SolAsym}) and the weak coupled case, (\ref{SolWAco0}) and (\ref{SolWAco1}), may suggest that the range of validity of this series expansion could be extended to larger values of $\beta$ in the perturbative approach.

\subsection{Near Horizon Solution}\label{NHSec}

Black holes must have an event horizon at $r_{+}$, where this radius is defined as the largest positive real root of $K(r_{+})=0$. Therefore it is useful to consider a series expansion of the metric function near the horizon given by
\begin{equation}\label{KHor}
K_{r_{+}}(r)=\sum_{n=1}^{\infty}{a_{n}\left(r-r_{+}\right)^{n}},
\end{equation}
which starts at $n=1$, because the metric function must vanish for $r=r_{+}$. Substituting this ansatz into the field equation (\ref{EQrr}) leads to a recurrence relations for the coefficients $a_{n}$. For example, the first recurrence relation is
\begin{eqnarray}\label{RReca}
&& 24\beta r_{+}^4a_{1}^2-4r_{+}^7a_{1}+4r_{+}^6-4g_{1}q^2r_{+}^4+g_{2}q^4=0,\nonumber
\end{eqnarray}
which may be solved for $a_{1}$, resulting in the following two solutions
\begin{equation}\label{Branchapm}
a_{1}^{\pm}=\frac{1}{12\beta {r_{+}}^2}\left[{r_{+}}^5\pm\sqrt{6\left(-g_{2}q^4+4g_{1}q^2{r_{+}}^4-4{r_{+}}^6\right)\beta+{r_{+}}^{10}}\right].
\end{equation}
Of these solutions only the $a_{1}^{-}$ branch recovers the limit of General Relativity when $\beta\rightarrow 0$. One way to reach this conclusion in detail is to expand the $a_{n}$ coefficients in powers of $\beta$, 
\begin{equation}\label{ASerie}
a_{n}=\sum_{j=0}^{j_{max}}h_{n,j}\beta^{j},
\end{equation}
and with the ansatz (\ref{KHor}) substitute into the field equation (\ref{EQrr}). The result is double-expansion series in $\beta$ and $(r-r_{+})$, whose first order term in $\beta$, $h_{n,0}$, correspond to the series expansion of the near horizon solution of the EH black hole of General Relativity, whose dominant contributions around the horizon are
\begin{eqnarray*}
&& h_{1,0}=\frac{g_{2}q^4-4g_{1}q^2 {r_{+}}^4+4 {r_{+}}^6}{4 {r_{+}}^7},\quad h_{2,0}=-\frac{g_{2}q^4-2g_{1}q^2 {r_{+}}^4+ {r_{+}}^6}{{r_{+}}^8}.
\end{eqnarray*}
However, one need to go beyond the leading order in $\beta$ to determine the correct coefficient for $a_{1}$. Actually, the next order in both expansions,
\begin{eqnarray*}
&& h_{1,1}=\frac{3 \left(g_{2}q^4-4g_{1}q^2 {r_{+}}^4+4 {r_{+}}^6\right)^2}{8 {r_{+}}^{17}},\quad h_{1,2}=\frac{9 \left(g_{2}q^4-4g_{1}q^2 {r_{+}}^4+4 {r_{+}}^6\right)^3}{8 {r_{+}}^{27}},\nonumber\\
\end{eqnarray*}
correspond to the leading terms in a $\beta$ expansion of (\ref{Branchapm}) for the negative sign. In summary, the correct coefficient is
\begin{equation}
a_{1}^{-}=\sum_{j=0}^{\infty}h_{1,j}\beta^{j}=\frac{1}{12\beta {r_{+}}^2}\left[{r_{+}}^5-\sqrt{6\left(-g_{2}q^4+4g_{1}q^2{r_{+}}^4-4{r_{+}}^6\right)\beta+{r_{+}}^{10}}\right].
\end{equation}
For completeness, we show the coefficients $h_{2j}$ and $h_{3j}$, for $a_{2}$ and $a_{3}$, with $j_{max=2}$ in Appendix \ref{NeaHor}.
Returning to the series (\ref{KHor}), the equation arising to order $\mathcal{O}((r-r_{+})^{2})$ is 
\begin{eqnarray}
&& 144a_{1}\beta r_{+}^5(2+a_{1}r_{+})a_{3}+96a_{1}\beta r_{+}^6a_{2}^2-8r_{+}^4(36a_{1}\beta+24a_{1}^2\beta r_{+}-r_{+}^3)a_{2}\nonumber\\
&& -8r_{+}^3(3r_{+}^2-4a_{1}r_{+}^3-12a_{1}^3r_{+}\beta - 2g_{1}q^2)=0.
\end{eqnarray}
This equation we can solve for $a_{3}$, leaving $a_{2}$ as a free parameter of the theory. If we consider the equation arising to order $\mathcal{O}((r-r_{+})^{3})$ the coefficient $a_{4}$ can be solved in terms of $a_{2}$.

\begin{figure}[h!]
\centering
\includegraphics[scale=0.95]{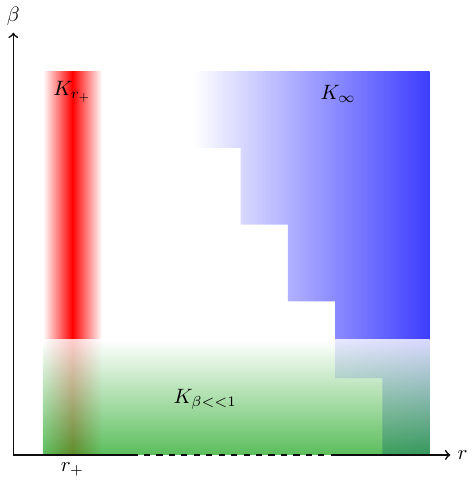}
\hfil  
\includegraphics[scale=0.95]{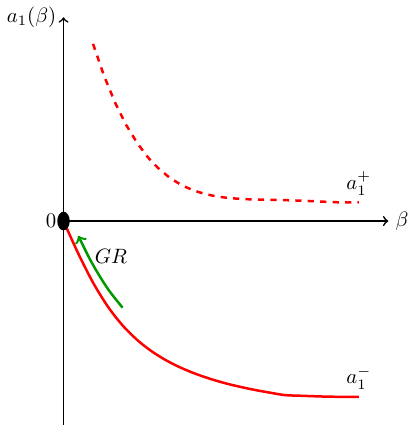}
\caption{Left: schematic range of validity of the three series expansion solutions (red $\rightarrow$ near horizon solution, blue $\rightarrow$ asymptotic solution, green $\rightarrow$ weak coupled solution). The straight cuts or steps represent the order (in $\beta$) at which the series expansion was constructed. Right: branches of the near horizon solution (\ref{Branchapm}). The Einsteinian branch is $a_{1}^{-}$, while the non-Einsteinian branch is $a_{1}^{+}$, which departs farther away from GR as $\beta\rightarrow 0$.}\label{DSol}
\end{figure}

The solutions (\ref{SolAsym}), (\ref{SolWAco0}) y (\ref{KHor}) are represented graphically in Figure \ref{DSol}, as well as the two branches arising in the near horizon solution. Although each solution is valid for different regimes, there are limits in which these solutions overlap, for example, if in the asymptotic and weak coupling solutions we consider the limits $\beta\rightarrow 0$ and $r\rightarrow\infty$, respectively these solutions behave in the same way. This suggests that there are valid solutions for the whole parameter space defining the theory.

\section{Properties of the solution}\label{Properties}

The purpose of this section is to study different physical properties of our solutions, using in particular the weak coupling solution (\ref{SolWAco0}) to linear order of the gravitational coupling constant $\beta$. First, we examine the behavior of the event horizon due to the presence of the cubic curvature terms and the nonlinear electrodynamics term as a function of electric charge. Subsequently, we analyze the effective potential that determines the motion of massless test particles, then describe in some detail how gravitational lensing is modified around these objects in the weak field limit and, finally, how additional curvature and electrodynamics terms in the action (\ref{Action}) modify the shadow of the black hole. 

\subsection{Effect of higher order temrs in the horizon}

\begin{center}
\begin{figure}[h!]
\includegraphics[scale=0.45]{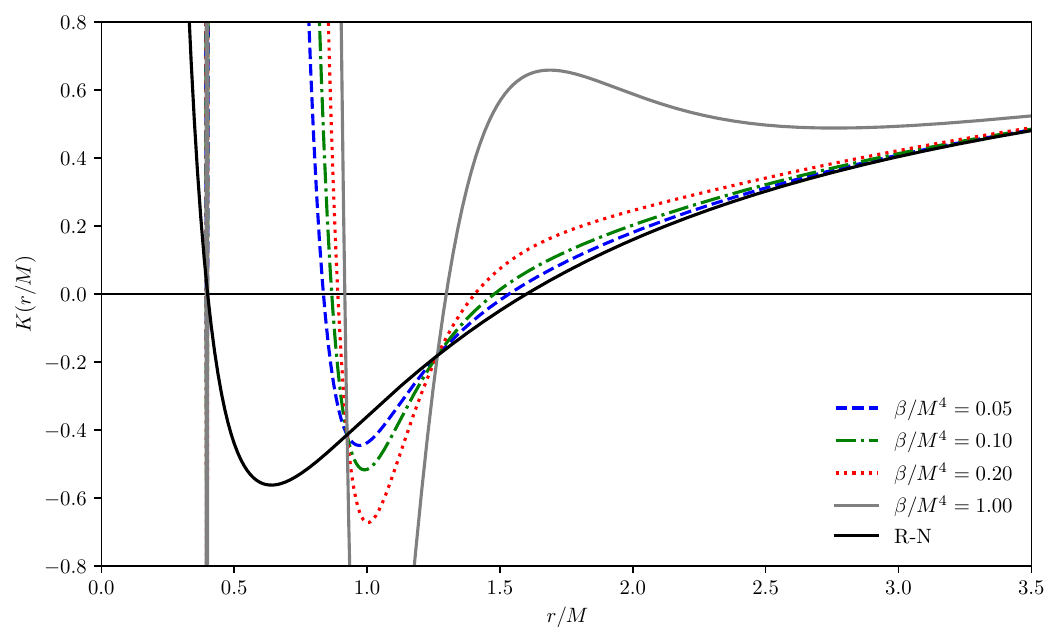}
\hfil
\includegraphics[scale=0.45]{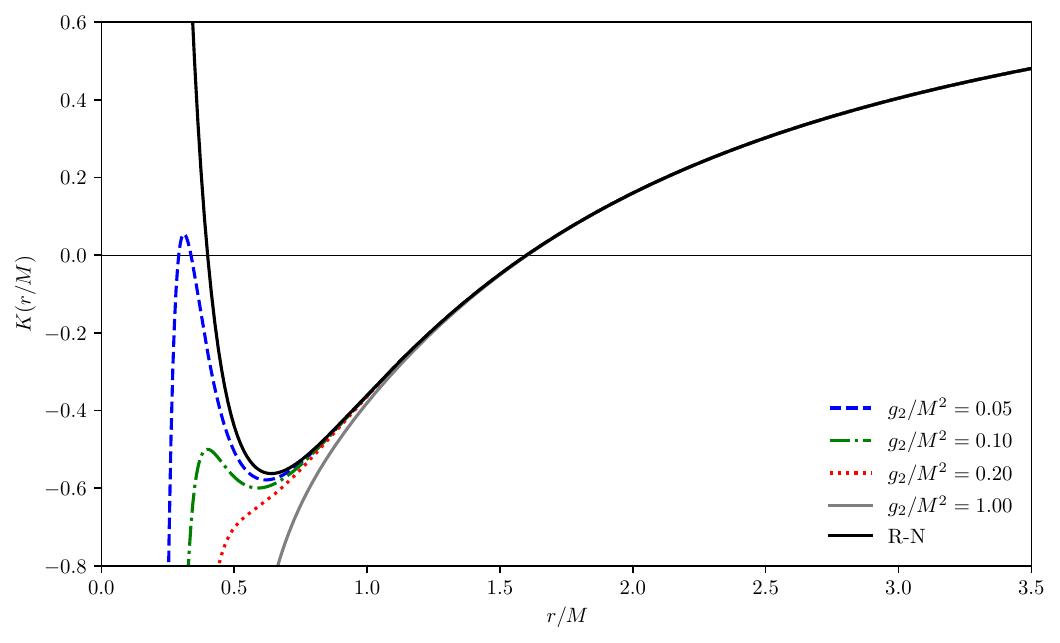}
\caption{Metric Function $K$ as a function of the radial distance from the center for different values of the cubic gravitaional coupling constant $\beta/M^{4}$ (left, with $g_{2}/M^{2}=0.00$ and $q/M=0.80$ fixed), or the non-linear electromagnetic parameter $g_{2}/M^{2}$ (right, with $\beta/M^{4}=0.00$ and $q/=0.80$ fixed). Both cases are compared to the Reissner-Nordström (R-N) function (black solid curve).}
\label{Ka1}
\end{figure}
\end{center}

As previously explained, the main characteristic of a black hole is the existence of an event horizon, whose location is determined by the largest zero of the metric function $K(r)$. Actually, the number of zeros is not preserved under changes of the theoretical parameters. As an example, Figure \ref{Ka1} shows the behaviour of $K(r)$ for different values of the coupling constants $\beta$ and $g_{2}$. The horizon moves to the smaller radii for larger values of gravitational coupling $\beta$, while it practically remains unaffected by changes in the EH parameter $g_2$. 

\begin{figure}[h!]
\centering
\includegraphics[scale=0.45]{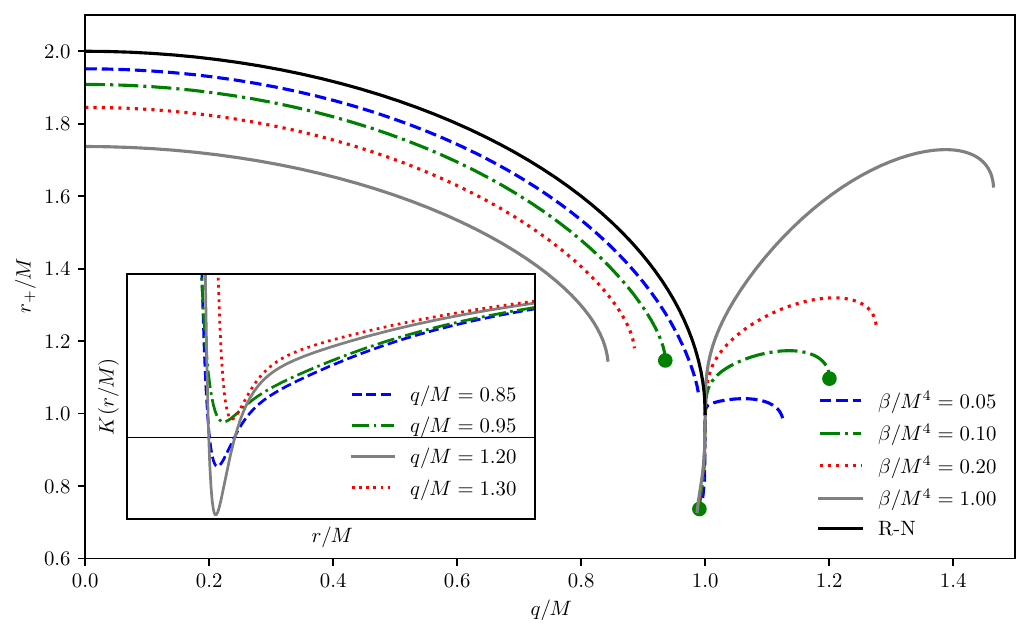}
\hfil
\includegraphics[scale=0.45]{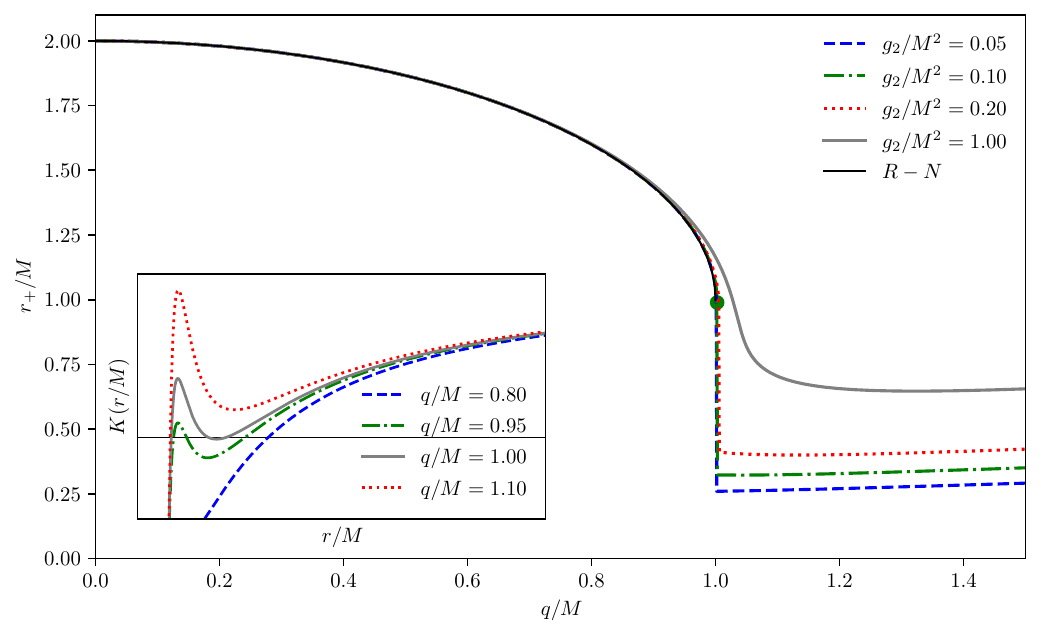}
\caption{Event horizon radius as a function of the electric charge using different values of the cubic gravitational coupling constant $\beta/M^{4}$ (left, with $g_{2}/M^{2}=0.00$), or the of the EH parameter $g_{2}/M^{4}$ (right, with $g_{2}/M^{4}=0.00$). The green circles represent extreme black holes, and are only shown for one particular value of the couplings ($\beta/M^{4}=0.10$ on the left and $g_{2}/M^{2}=0.10$ on the right). The Reissner-Norsdtröm (R-N) black hole is the black solid line. The insets show the corresponding the metric function $K(r)$ for different values of the electric charge with $\beta/M^{4}=0.10$ (left) or $g_{2}/M^{2}=0.10$ (right).}
\label{RHQBeta}
\end{figure}

Alternatively, in Figure (\ref{RHQBeta}) we show the value of the largest root of $K(r)$ as a function of the theory parameters. As appreciated in the left plot, increasing the value of the coupling constant $\beta$ reduces the size of the event horizon but it is also observed that the event horizon reduces its size when the value of the electric charge increases and this occurs similarly to the case of the Reissner-Nordstrom black hole (R-N) up to a critical value which is where the case of an extreme black hole exists, It is also interesting to note that from this critical point there is a range of values of the electric charge in which there is no black hole, but subsequently these solutions would be allowed with a larger electric charge than that of R-N.

\begin{figure}[h!]
\centering
\includegraphics[scale=0.45]{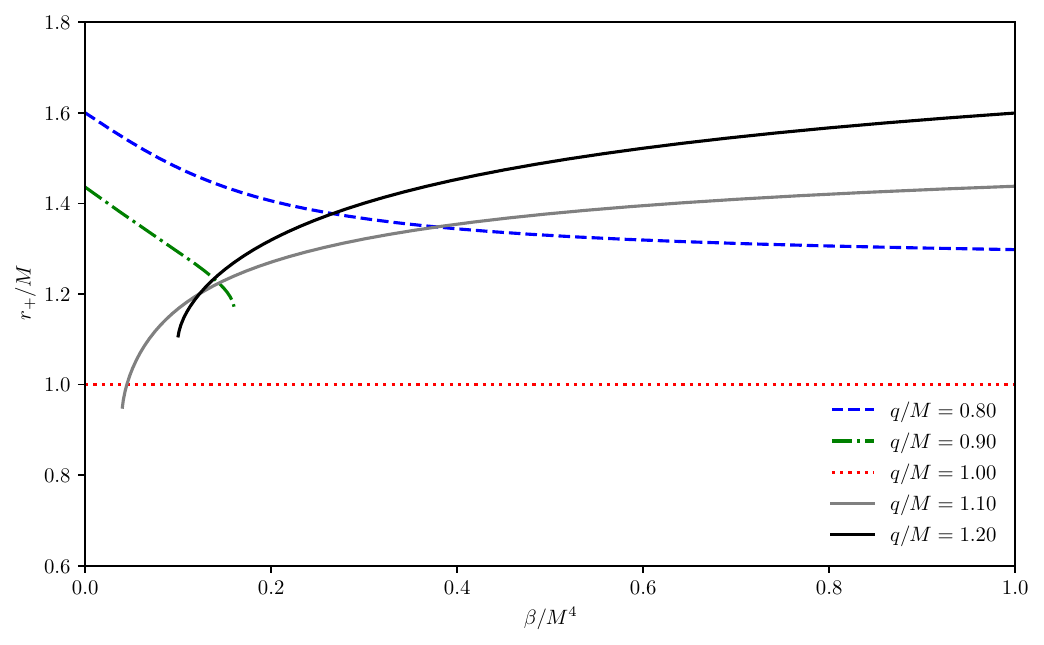}
\hfil
\includegraphics[scale=0.45]{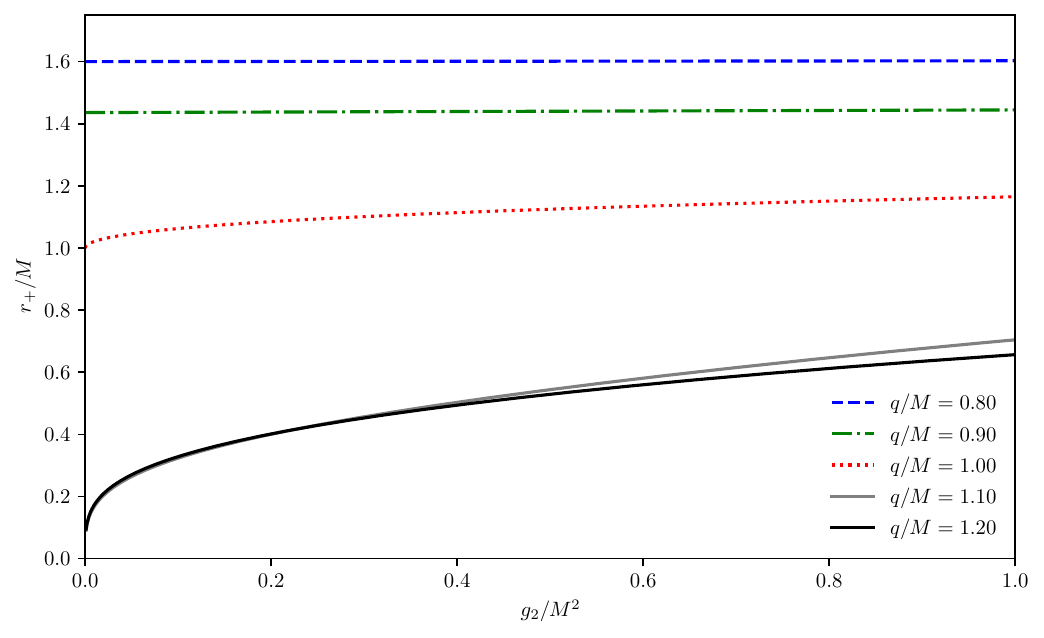}
\caption{Radius of the event horizon as a function of the gravitational coupling constant $\beta/M^{4}$ (left) or the EH parameter $g_{2}/M^{2}$ (right) for different values of the electric charge, with the other higher order term coupling switched off. Extreme black hole cases are located where the curves begin or end.}
\label{RHG2}
\end{figure}

One further combination of parameters is to switch off one only one the higher order terms and then vary the electric charge. This is shown in Figure (\ref{RHG2}), where we notice that the radius of the event horizon remains constant as a function of the cubic gravitational coupling when the electric charge has the extreme Reissner-Norsdtröm value, $q/M=1.00$, but then it varies interestingly otherwise. In the plot on the right the radius of the event horizon is written as a function of the EH parameter with $\beta/M^{4}=0.00$, we observe that the event horizon begins to present differences with respect to the Reissner-Norsdtröm case when the electric charge is $q/M=1.00$, this curve would represent the points where the gap appears in the event horizon. 

\subsection{Motion of massless test particles}

\begin{center}
\begin{figure}[h!]
\includegraphics[scale=0.45]{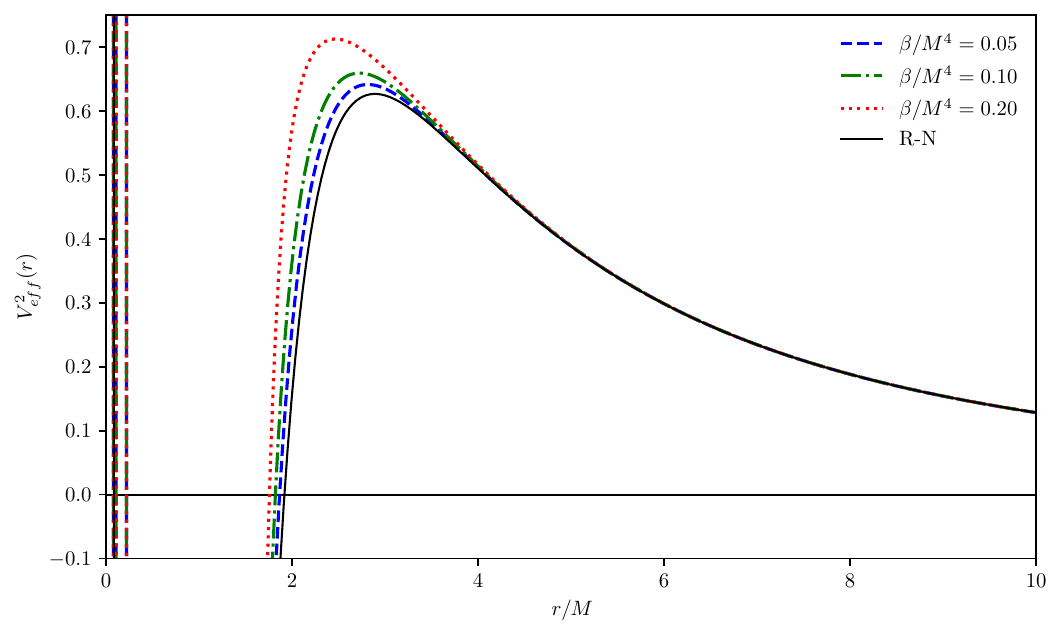}
\hfil
\includegraphics[scale=0.45]{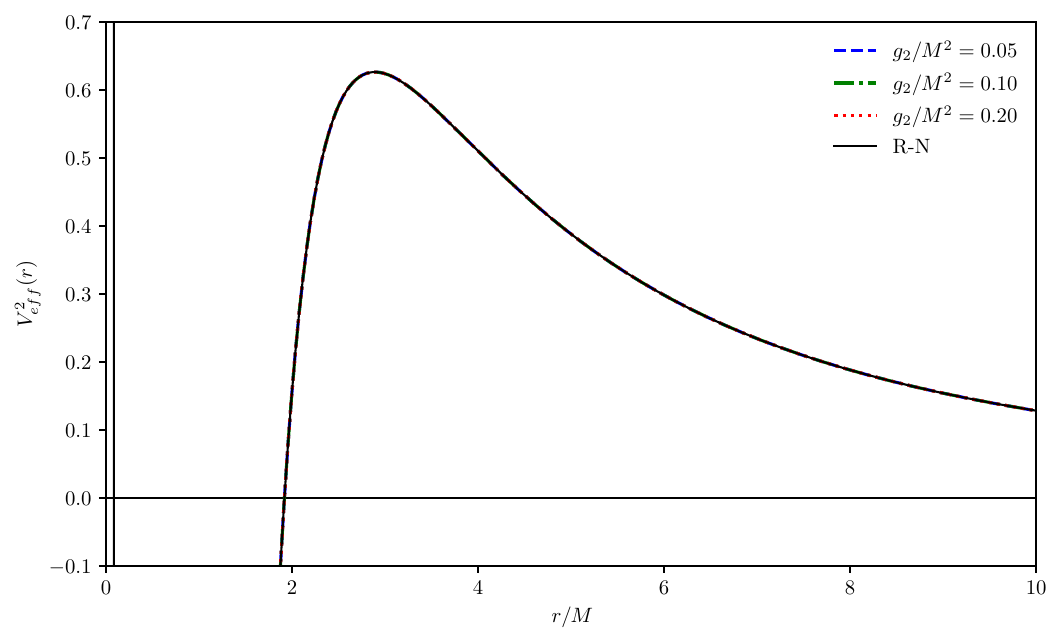}
\caption{Effective potential for massless particles as a function of radial coordinate, with $q/M=0.40$ and $l/M=4.00$, for different values of $\beta$ (left, with $g2=0$), or $g_2$ (right, with $\beta=0$). Both two cases are contrasted against the Reissner-Nordström case (black solid curve).}
\label{EffPot}
\end{figure}
\end{center}

As in the case of General Relativity due to the spherical symmetry of the metric (\ref{metric}) one can study the motion of massless test particles in the equatorial plane ($\theta=\pi/2$) for the solutions obtained in the previous section. In general, for metrics of the form (\ref{metric}), the lagrangian for test particles can be written as
\begin{equation}\label{LagTP}
\mathcal{L}=\frac{1}{2}\left(-K(r)\dot{t}^2+\frac{\dot{r}^2}{K(r)}+r^2\dot{\phi}^2\right),
\end{equation}
where $\dot{x}^{\alpha}=dx^{\alpha}/d\lambda$, with $\lambda$ as an affine parameter for a massless particle. It is well known that one can associate conserved quantities to coordinates that do not appear explicitly in a Lagrangian. The previous Lagrangian does not explicitly depend on the coordinates $t$ and $\phi$, which result in the energy ($\mathcal{E}$) and angular momentum ($l$) of the test particle being conserved due to the Nöther's theorem. Using the norm of the vector $g_{\alpha\beta}{\dot{x}}^{\alpha}{\dot{x}}^{\beta}=0$ the equation of motion of the coordinate $r$ can be written as 
\begin{equation}
\dot{r}^2=\mathcal{E}^2-\frac{l^2}{r^2}K(r).
\end{equation}
Writing it as $\dot{r}^2=\mathcal{E}^2-V_{eff}^2(r)$, we can identify the effective potential as
\begin{equation}\label{PotEfec}
V_{eff}^{2}(r)=\frac{l^2}{r^2}K(r),
\end{equation}
which governs the radial motion of the test particles and defines the type of orbits through which they can move.  In Figure (\ref{EffPot}), we show the dependence of the effective potential on the higher order term couplings $\beta/M^{4}$ and $g_{2}/M^{2}$. The parameter $\beta$ causes the maximum of the potential to grow and to be located closer to the event horizon, while the EH parameter does not play any role, hence $V_{eff}$ is the same as that for Reissner-Nordström case.
The presence of a maximum in the potential implies the existence of unstable circular orbits (UCO); which is related to the photonsphere, and must satisfy the condition
\begin{subequations}
   \begin{equation}\label{CUSCOa}
       V_{eff}^2(r_{c})=\mathcal{E},
   \end{equation}
   \begin{equation}\label{CUSCOb}
       V_{eff}^2(r_{c})'=0.
   \end{equation}
\end{subequations}
From this condition we obtain the radius of the photonsphere, also known as the last photon orbit, while from the condition (\ref{CUSCOa}) we obtain the impact parameter, $\eta_{c}$. In summary, for static and spherically symmetric metrics, such as in our solutions, the resulting impact parameter is 
\begin{equation}\label{ImpPar}
\eta_{c}^{2}=\frac{l^2}{\mathcal{E}^2}=\frac{r_{c}^2}{K(r_{c})},
\end{equation}
where $r_{c}$ is a solution of the polynomial arising from the condition (\ref{CUSCOb}), which corresponds to the radius of the UCO for a massless particle. Now we are in a position to construct black shadows using the massless particle motion.

\subsection{Black hole shadow}

As we have just discussed, there is a possibility of finding the motion of massless test particles over circular orbits closed to the black holes. Due to its unstable nature, a perturbation in the particle dynamics would lead to a infall trajectory into the black hole or to escape to infinity. Those particles that avade the black hole can be captured by observatories on Earth, with a deformed ring as a spatial distribution. This observations is the so-called shadow of the black hole, and so far we have only detected the shadow of the supermassive black holes M87\cite{Akiyama1} and SgrA$^*$\cite{Akiyama3} by the Event Horizon Telescope collaboration (EHT). 

\begin{figure}[h!]
\centering
\includegraphics[scale=0.45]{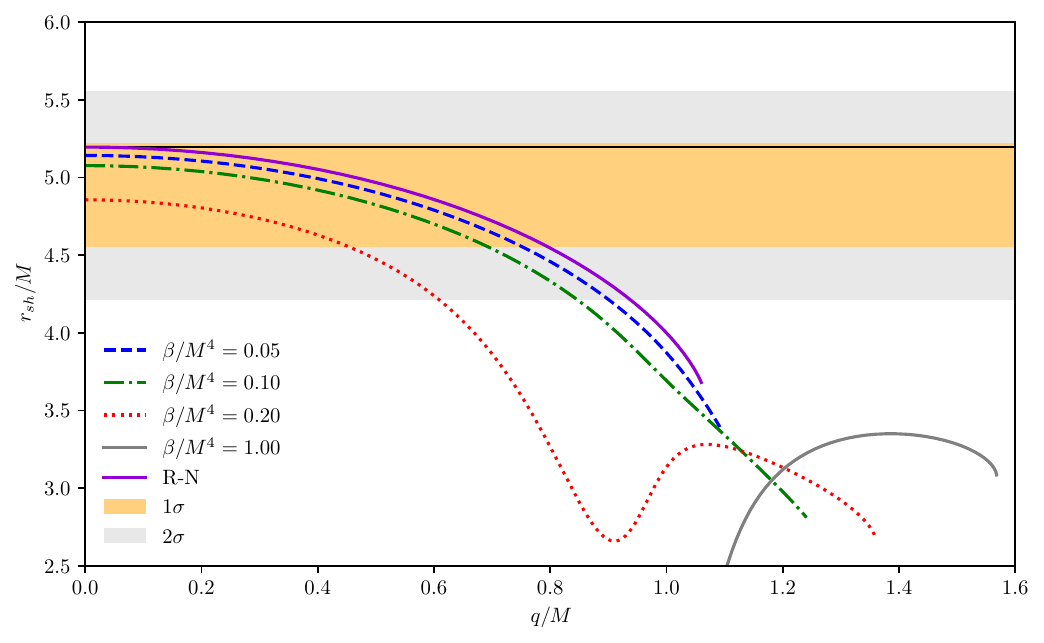}
\hfil
\includegraphics[scale=0.45]{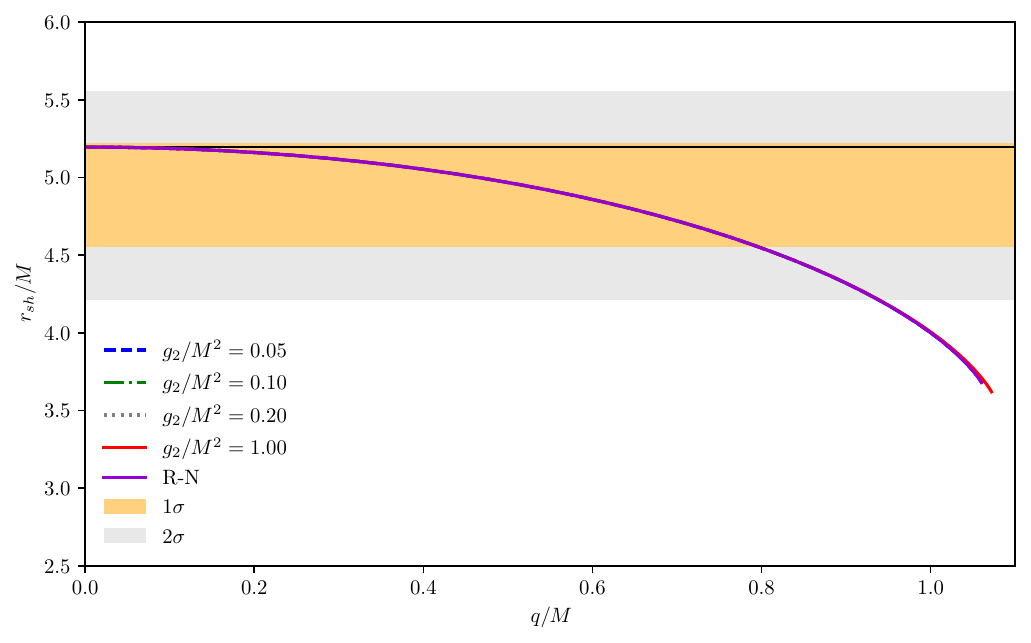}        
\caption{Left: the radius of the shadow of the charged black hole in ECG as a function of electric charge for different values of the coupling constant $\beta/M^{4}$ with $g_{2}/M^{2}=0.00$. Right: the radius of the shadow as a function of electric charge for different values of the EH parameter $g_{2}/M^{2}$ with $\beta/M^{4}=0.00$. In both cases the horizontal black line represents the shadow radius of the Schwarzschild black hole which is used as a reference for SgrA*.}
\label{ShadowQBeta}
\end{figure}

\begin{figure}[h!]
\centering
\includegraphics[scale=0.45]{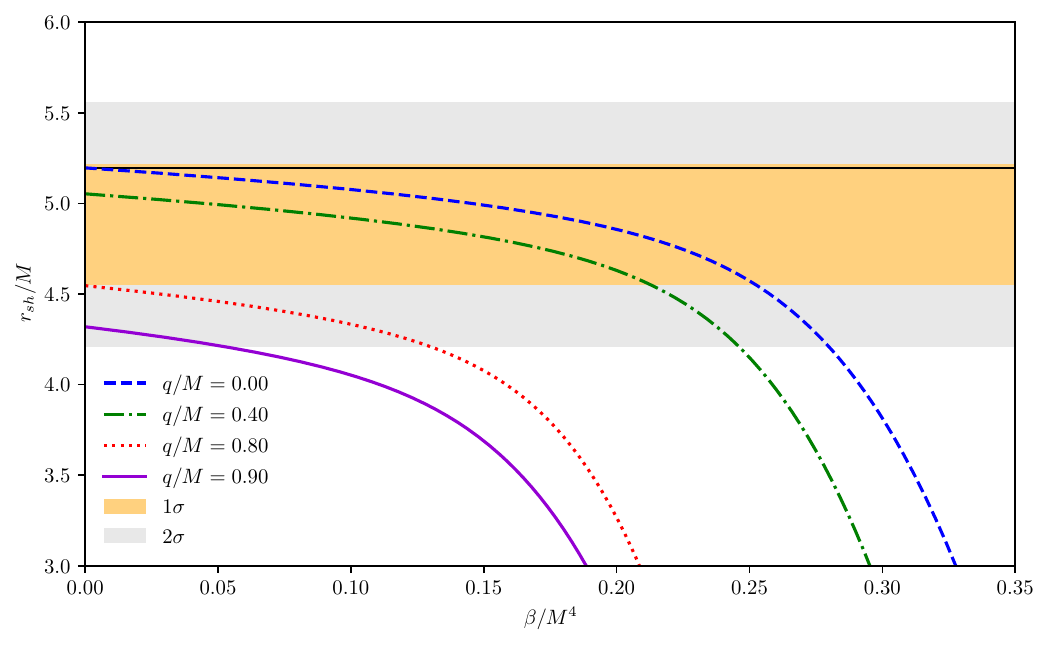}
\hfil
\includegraphics[scale=0.45]{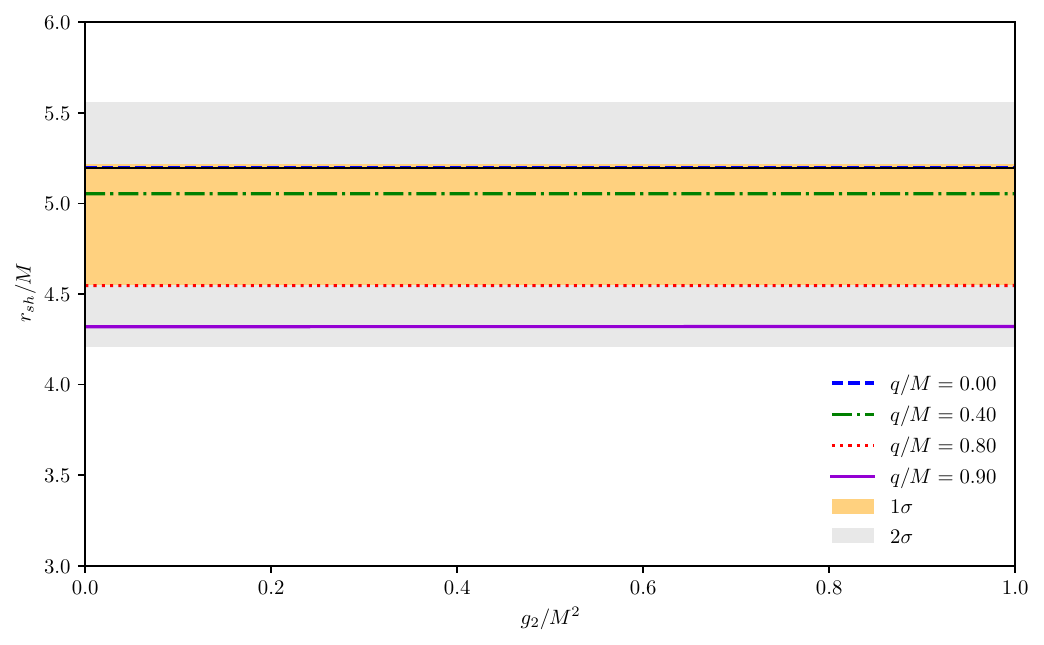}
\caption{Left: the radius of the shadow of the charged black hole in ECG as a function of the coupling constant $\beta/M^{4}$ for different values of the electric charge with $g_{2}/M^{2}=0.00$. Right: the radius of the shadow of the charged black hole in ECG as a function of the EH parameter $g_{2}/M^{2}$ for different values of the electric charge with $\beta/M^{4}=0.00$. In both cases the horizontal black line represents the shadow radius of the Schwarzschild black hole which is used as a reference for SgrA*.}
    \label{ShadowQG2}
\end{figure}

To determine the radius of the shadow's contour one may use the optical approximation where this radius of the photosphere, $r_{sh}$, is approximated by the critical impact parameter, $\eta_{c}$ \cite{Perlick}. From equation (\ref{ImpPar}), one obtains
\begin{equation}\label{RSh}
    r_{sh}^{2} = \eta_{c}^{2} = \frac{r_{c}^{2}}{K(r_{c})}.
\end{equation}
One may use observational data to constrain the theory parameters by studying how they affect the black hole radius shadow. We follow the approach \cite{Vagnozzi} in order to compare this shadow radius with the SgrA* observations, in which the uncertainties allowed by the EHT are considered. In this cited work, the fractional deviation $\delta$ between the black hole shadow radius $r_{sh}$ and its Schwarzschild counterpart (defined by $\delta  = r_{sh}/(3M\sqrt{3})-1$) is constraint to $-0.125\gtrsim\delta\gtrsim 0.005$ or $-0.19\gtrsim\delta\gtrsim 0.07$ when considering a $1\sigma$ or $2\sigma$ trust intervals respectively. We can use these constraints to place an upper bound for the electric. For the Reissner-Norsdtröm black hole (RN) and up to two sigma, the electric charge cannot be larger than 80\% the value of the mass. However, as we increase the cubic gravitational coupling, this upper bound decreases (see Figure \ref{ShadowQBeta}). In contrast, the nonlinear electrodynamical term does not modify the shadow radius as a function of electric charge, hence recovering the RN prediction. The same observational data can be exploited to place constraints in the coupling constants of the cubic curvature and quadratic electrodynamics terms. As also seen for the electric charge, the shadow radius also decreases as a function of the cubic gravity coupling, $\beta$, but remains unaffected by the value of the quadratic electrodynamics term, $g_2$. Moreover, after an initial constant decay of the radius as a function of $\beta$ there is a turning point where the suppression gets stronger, which is better appreciated in Figure (\ref{ShadowQG2}). In summary, there is an effect on the black hole shadow as a function of the different couplings which can be used to contraint the model when trying to fit the ETH data. However, one should bare in mind that these observations really take place on the Earth's detectors, so one should bring the black hole shadow to the observer, where a good approximation is to place her at infinity. 

\subsubsection{Observer at infinity}

\begin{center}
\begin{figure}[h!]
\includegraphics[scale=0.45]{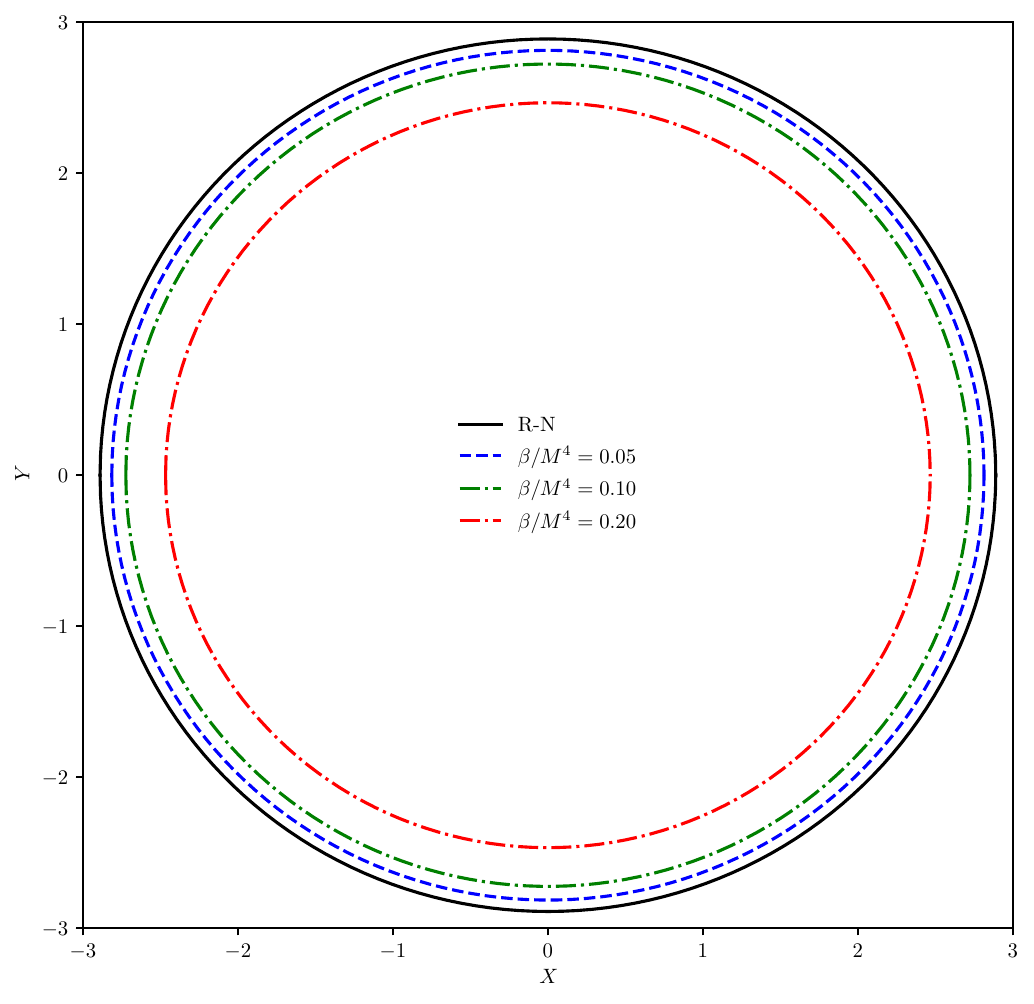}
\caption{Contour of the black hole shadow as seen by an observer at infinity. Increasing the cubib gravity coupling, $\beta/M^{4}$, leads to a smaller shadow's radius when compared to the Reissner-Norsdtröm (RN) black hole case. We consider that $q/M=0.40$ and $g_{2}/M^{2}=0.00$.}
\label{Contour}
\end{figure}
\end{center}

For asymptotically flat spacetimes, the radius of the shadow measured by an observer located at infinity is determined by means of the celestial coordinates X and Y, which define a plane of projection on the celestial sphere, defined as

\begin{equation}
X=\lim_{r_{0}\rightarrow\infty}\left(-r_{0}^{2}\sin{\theta_{0}}\left.\frac{d\phi}{dr}\right|_{\theta=\theta_{0}}\right),\quad Y=\lim_{r_{0}\rightarrow\infty}\left(r_{0}^2\left.\frac{d\theta}{dr}\right|_{\theta=\theta_{0}}\right),
\end{equation}
where $r_{0}$ is the distance from the observer to the black hole and $\theta_{0}$ is the angle of inclination between the line of sight of the observer and the rotation axis of the black hole. As in the case of the motion of the test particles, due to the spherical symmetry of the solutions obtained, we can consider, without loss of generality, that the observer is located in the equatorial plane $\theta=\pi/2$, therefore the contour of the shadow of the black hole is given by 

\begin{equation}\label{CShadow}
X^2 + Y^2 = \eta_{c}^{2},
\end{equation}
from this expression we can confirm that the radius of the shadow corresponds to the critical impact parameter evaluated at the radius of the photosphere, as we had already indicated in eq. (\ref{RSh}). Figure (\ref{Contour}) illustrates how the observed shadow radius decreases as the coupling constant $\beta/M^{4}$ grows. The case where the value of the $g_2$ parameter varies is not illustrated, since it does not affect appreciably the behavior of the event horizon and the effective potential, (see Figures (\ref{RHQBeta}) and (\ref{EffPot})), hence nor the observed shadow. 

\subsection{Gravitational lensing in the weak field limit}

One of the first effects used to test the theory of General Relativity was gravitational lensing, which consists of the deviation of the trajectory of light when passing near a massive object. Recently, these lensing effects has been used to constrain the compactness and distortion parameters of gravitational lenses; e.g. when applied to the supermassive black holes M87$^*$ and SgrA$^*$ in \cite{Virbhadra1,Virbhadra2}. When considering the weak field limit, the lensing effect around our black holes occurs when massless particles pass far away from the photosphere and are slightly deflected from their trajectory. In this case, the deflection angles are assumed to be small, thus can be determined by the Gaussian curvature and the Gauss-Bonnet theorem for the optical metric using the method developed by Gibbons and Werner\cite{GibbonsWL} (for further details see example \cite{OkyayWL}). 

The optical metric for null geodesic $ds^{2}=0$ using the static, spherically symmetric metric  of (\ref{metric}), results in the equality   
\begin{equation}\label{optic}
    dt^2=\gamma_{ij}^{opt}dx^{i}dx^{j}=\frac{dr^2}{(K(r))^{2}}+\frac{r^{2}}{K(r)}\left(d\theta^2+\sin{(\theta)}^2\right)d\phi^{2}.
\end{equation}
If again we consider the motion along the equatorial plane $\theta=\pi/2$, the Gaussian curvature, which is defined as $\mathcal{K}=R/2$ (with $R$ being the Ricci scalar for the optical metric (\ref{optic})), is given by 
\begin{equation}
    \mathcal{K}= \frac{1}{2}\left(\frac{d^{2}}{dr^{2}}K(r)\right)K(r)-\frac{1}{4}\left(\frac{d}{dr}K(r)\right)^{2};
\end{equation}
which by applying the Gauss-Bonnet theorem, that connects the intrinsic differential geometry of a surface to its topology, leads to an expression for a small variation of the deflection angle $\delta\alpha$ \cite{GibbonsWL}, namely
\begin{equation*}
    \delta\alpha = -\int\int_{D} \mathcal{K}dS,
\end{equation*}
where $dS=\sqrt{-\gamma}rdrd\phi$, with $\gamma$ being the determinant of the optical metric. The limits of integration are $\frac{\eta}{\sin{(\theta)}}<r<\infty$ and $0<\phi<\pi$, where the impact parameter is $\eta$. The deflection angle can also be written as a power series in $\beta$, which to order $\mathcal{O}(\beta)$ is given by
\begin{equation}
    \delta\alpha = \delta\alpha_{0}+\beta\delta\alpha_{1}+\cdots\nonumber\\, 
\end{equation}
where the coefficients $\delta\alpha_{i}$ are 
\begin{eqnarray}    
\delta\alpha_{0}&& = \frac{4M}{\eta}-\frac{3\pi q^{2}}{4\eta^{2}}+\frac{7g_{2}\pi q^{4}}{128\eta^{6}}-\frac{3M^{2}\pi}{4\eta^{2}}+\frac{8Mq^{2}}{3\eta^{3}}-\frac{3\pi q^{4}}{16\eta^{4}}-\frac{304g_{2}Mq^{4}}{1225\eta^{7}}+\frac{63g_{2}\pi q^{6}}{2048\eta^{8}}+\frac{231g_{2}^{2}\pi q^{8}}{409600\eta^{12}},  \nonumber\\
\delta\alpha_{1}&& = -\frac{945 M^2 \pi }{4 \eta ^6}+\frac{512 M (1157 M^2 + 756 q^2)}{245 \eta^7}-\frac{105 \pi (391 M^4 + 1224 M^2 q^2 + 144 q^4)}{64 \eta^8}+\frac{8192 M q^2 (719 M^2 + 555 q^2)}{945 \eta^9}\nonumber\\
&& -\frac{63 \pi q^4 (10506 M^2 + 1789 q^2)}{320 \eta^{10}}+\frac{8192 M q^4 (395 q^2 - 63 g_{2})}{1155 \eta^{11}}+\frac{231 \pi q^4 (2859 g_2 M^2 + 624 g_2 q^2 - 760 q^4)}{1280 \eta^{12}}\nonumber\\
&& -\frac{8192 g_2 M q^4 (35246 M^2 + 49107 q^2)}{195195 \eta^{13}}+\frac{3861 g_2 \pi q^6 (15338 M^2 + 3625 q^2)}{71680 \eta^{14}}-\frac{78675968 g_{2} M q^8}{53625 \eta ^{15}}\nonumber\\
&& +\frac{1287 g_2 \pi q^8 (10945 q^2 - 1224 g_2)}{163840 \eta^{16}}+\frac{711131136 g_2^2 M q^8}{5165875 \eta ^{17}}-\frac{7293 g_2^2 \pi q^8 (4173 q^2 + 8672 M^2)}{1310720 \eta^{18}}\nonumber\\
&& +\frac{17376608256 g_{2}^2 M q^{10}}{109698875 \eta ^{19}}-\frac{53625429 g_{2}^2 \pi  q^{12}}{4096000 \eta ^{20}}+\frac{7386137577 g_{2}^3 \pi  q^{12}}{11534336000 \eta ^{22}}-\frac{8167882752 g_{2}^3 M q^{12}}{1943612125 \eta ^{23}}+\frac{2845448151 g_{2}^3 \pi  q^{14}}{4194304000 \eta
^{24}}\nonumber\\
&& -\frac{539768853 g_{2}^4 \pi  q^{16}}{46976204800 \eta ^{28}}.\nonumber
\end{eqnarray}

The coefficient $\delta\alpha_{0}$ corresponds to the result of General Relativity. The firs three terms had already been considered in \cite{QMFu}. Notice that the corrections to the deflection angle due to the nonlinear electrodynamics term influence up to order $\mathcal{O}\left(\eta^{-12}\right)$. In addition, from $\delta\alpha_{1}$ we appreciate that the deflection angle correction due to the cubic gravity appears at order $\mathcal{O}\left(\eta^{-6}\right)$, while the combined correction, electromagnetic and gravity, arises first at order $\mathcal{O}\left(\eta^{-11}\right)$.

\begin{figure}[h!]
\centering
\includegraphics[scale=0.45]{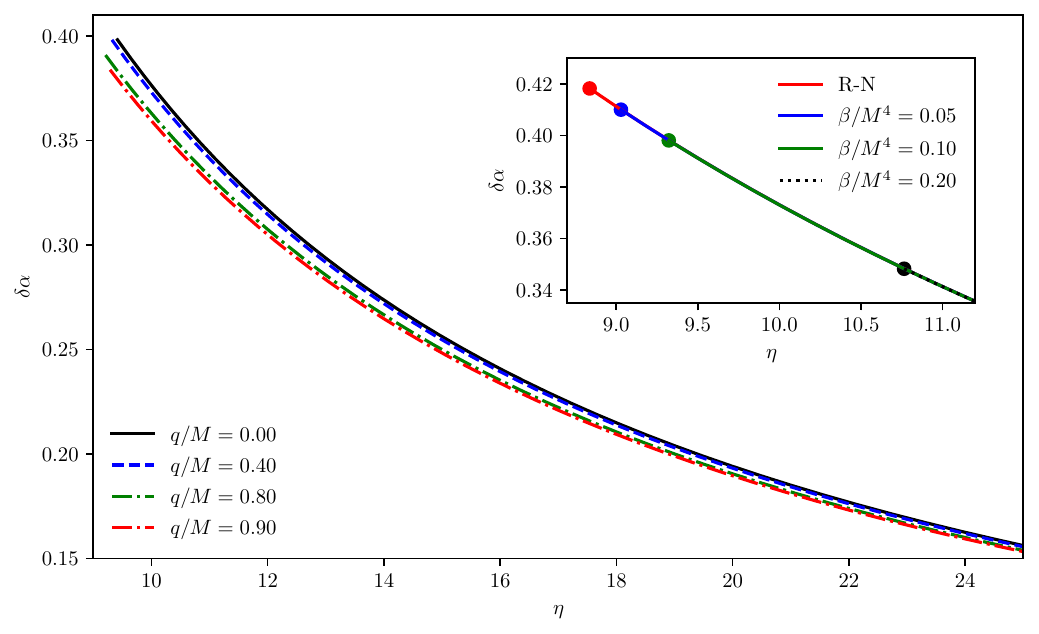}
\hfil
\includegraphics[scale=0.45]{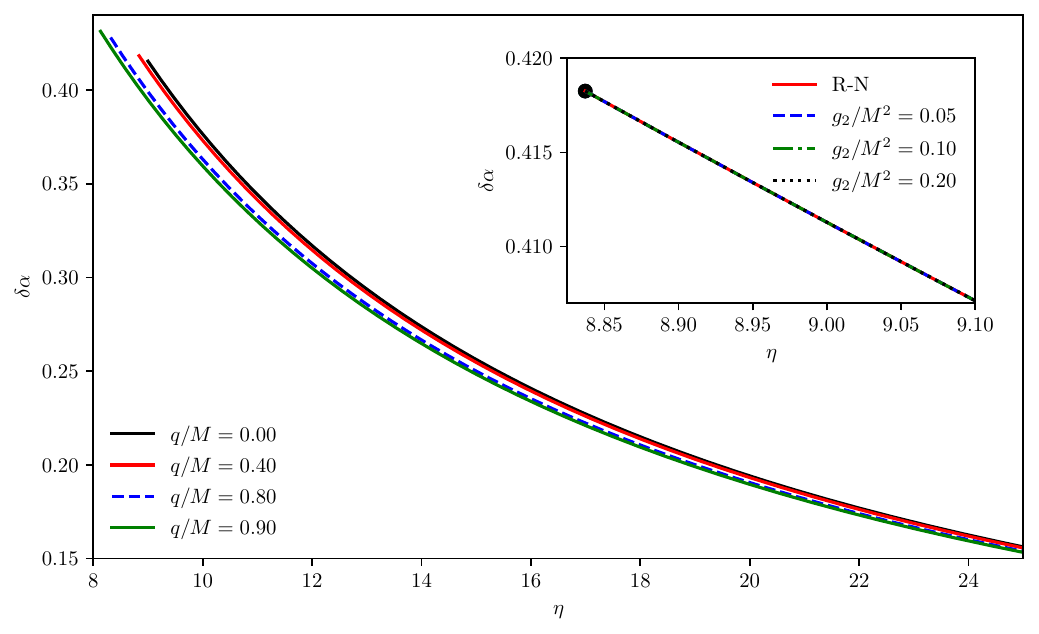}
\caption{Variation of the lensing deflection angle with respecto to the impact parameter, for different values of electrical charge. Left: the case with $\beta/M^{4}=0.10$ and $g_{2}/M^{2}=0.00$. The inner box shows the variation with respect to the  gravitational coupling $\beta/M^{4}$, instead, and the dots indicate the location of the critical impact parameter for each value of $\beta/M^{4}$. Right: the case with $g_{2}/M^{2}=0.10$ and $\beta/M^{4}=0.00$. The inner box shows the behavior with the $g_{2}/M^{2}$ coupling and the dots indicate the location of the critical impact parameter for each value of the EH parameter, where no change is observed with respect to the Reissner-Norsdtröm (RN) case. On both sides, the black curve represents the neutral black hole case for $\beta/M^{4}\neq 0.00$ and $g_{2}/M^{2}\neq0.00$, respectively. The curves start at the critical impact parameter $\eta_{c}$ associated with each value of the electric charge $q/M$. In all cases $\delta\alpha\rightarrow 0$ when we consider the limit $\eta\rightarrow\infty$.}
\label{AngDef}
\end{figure}

Figure \ref{AngDef} shows the variation of the deflection angle as a function of the impact parameter for different values of the electric charge. We note that by increasing the electric charge the critical impact parameter is lower with respect to the neutral case and reduces the deflection angle. While variations of the constant $\beta$ produce an increase in the value of the critical impact parameter but maintaining the same deflection angle with respect to the R-N case.

\section{Final Comments}\label{Fcomm}

Effective field theories (EFT) are a successful approach to study nature, because not only they describe and predict phenomena, but also can be used to push an already accurate description of physical phenomena beyond their regime of validity. We follow this idea of a bottom up approach to extend General Relativity and Electromagnetism to higher energies, where we expect corrections that may lead to observable signatures in the physics of compact objects. Based on this working hypothesis, the Maxwell Lagrangian would be corrected by a tower of terms which involve higher powers of the spin one field, where the leading order would be a quadratic correction of the electromagnetic field strength. An equivalent expansion in gravity leads to corrections with higher powers of the curvature tensors. At this level the number of possibilities vastly grow but under certain conditions, such as demanding that Schwarszchild-like solutions exist, the leading correction to the Hilbert-Einstein Lagrangian is a cubic curvature term \cite{BuenoCano1}. 

On the other hand, the recently captured black hole shadows of SgrA$^*$ and M87$^*$ by the Event Horizon Telescope (EHT) may represent a nice testbed for these ideas, where the higher order corrections terms could have a visible signal. To that extent, we search for black hole solutions and describe how their idealised shadow radius behaves as a function of the theory parameters. We consider that these higher order corrections are small in both sectors (electromagnetic and gravitational), ensuring the validity of the effective field theory approach. However, we find non-negligible corrections if  the gravitational coupling is of the order of the black hole's mass. Therefore, one may wonder if this is small enough for the EFT to hold. If one considers a series of the form $\mathcal{R}+\beta\mathcal{R}^3$, where by curly $\mathcal{R}$ we schematically represent any scalar constructed by curvature tensors of the same order of derivatives as the Ricciscalar, then we can give a rough estimate of the $\beta$ coupling for the second term to be smaller than the first. In the case of of the  Schwarzschild metric, $\mathcal{R}=\sqrt{R_{abcd}R^{abcd}}=\frac{\sqrt{48}M}{r^{3}}$, while the curvature for SgrA* is of the order of $10^{-22}m^{-2}$\cite{Baker}. We find a bound for the gravitational coupling of $\beta\lesssim 0.26M^{4}$, which implies a cubic correction term of  $\beta\mathcal{R}^{2}\sim 3.659\times10^{-4}$. In summary, even though the gravitational coupling constant appears to be very large because $\beta\sim M^{4}$, the whole term remains small, hence within the EFT framework.

We use three perturbative expansions to construct solutions similar to the Schwarzschild or Reissner–Nordström metrics, which we know are exact solutions to the leading order terms in the EFT. Two of the expansions approximate the solution near the origin or asymptotically, while the third considers a series in the gravitational coupling $\beta$ (see Figure \ref{DSol}). The latter expansion matches exactly that of the asymptotic or near-horizon solutions in their corresponding radial limits, suggesting the $\beta$-series is valid beyond the perturbative hypothesis. Therefore, the solution with $\beta\geq 1$ which would also connect the near horizon and asymptotic regions could exist, but one would expect the presence of instabilities or ghosts on reasonable scales. However, by assuming the higher order corrections are small, those instabilities if present, should not be captured by physics outside the black hole horizon. Under this valid EFT approach, we constrain the model parameters using the black hole shadow analysis of the EHT on SgrA$^*$, finding larger departures from Reissner–Nordström or Schwarzschild due to the gravitational higher order term than the non-linear electrodynamics. Moreover, that strong dependence on the cubic gravity term is also present in the black hole horizon structure, where one can even find more than one value of the theory parameters to get extremal black holes.

Finally, the great EHT observations are not the only way to discard modifications to  gravity in our compact objects. There are also the impressive measurements of the gravitational waves detected by the LIGO/VIRGO collaboration\cite{LIGO}. Determining the quasinormal modes of our solutions would be a starting point to use these exquisite gravitational wave data, but we leave the corresponding effort to a future line of study, since it is beyond the objective of the present work.

\acknowledgments

GGC would like to thank Armando A. Roque for help with
Mathematica. GGC is also grateful to CONAHCYT for a Estancias Posdoctorales por M\'exico 2022 Grant No. 2668595.


\appendix 

\section{Corrections for weak coupling behavior.}

For the weak coupling solution, in the asymptotic case and in the near-horizon case, we consider power series of the coupling constant $\beta$. In Sections \ref{WCSec} and \ref{NHSec} we only showed the first terms of these series, now we propose to show the form of the terms corresponding to order $\mathcal{O}\left(\beta^{2}\right)$. 

\subsection{Small Coupling Constant $\beta$}\label{SmCoup}

For the weak coupling solution, we consider a series expansion in the coupling constant $\beta$ of the form  

\begin{equation*}
    K_{\beta<<1}(r)=\sum_{n=0}^{\infty}\beta^{n}k_{n}(r),
\end{equation*}

\noindent at each order of $\beta$ we have a first order differential equation. The solutions for $k_{0}(r)$ and $k_{1}(r)$ corresponding to the case of General Relativity and the correction to first order in $\beta$ are given by expressions (\ref{SolWAco}) and (\ref{SolWAco1}), respectively. The second order correction has the following form

\begin{eqnarray}
k_{2}(r)&&=-\frac{435456M^3}{r^{11}}+\frac{1772928M^4}{r^{12}}-\frac{1792896M^5}{r^{13}}+g_{1}q^2\left(\frac{1451520M^2}{r^{12}}-\frac{6801408M^3}{r^{13}}+\frac{7776384M^4}{r^{14}}\right)\nonumber\\
&& -g_{1}^2q^4\left(\frac{1461888M}{r^{13}}-\frac{8928576M^2}{r^{14}}+\frac{12476544M^3}{r^{15}}\right)+g_{1}^3q^6\left(\frac{460800}{r^{14}}-\frac{4874112M}{r^{15}}+\frac{9452544M^2}{r^{16}}\right)\nonumber\\
&& +g_{1}^4q^8\left(\frac{944640}{r^{16}}-\frac{3422976M}{r^17}\right)+\frac{g_{1}^5q^{10}477888}{r^{18}}-g_{2}\left[\frac{3079296M^2q^4}{5r^{16}}-\frac{864\left(14691M^2+5980g_{1}q^2\right)Mq^4}{5r^{17}}\right.\nonumber\\
&&\left. +\frac{96\left(135715M^4+291483M^2g_{1}q^2+22176g_{1}^2q^4\right)}{5r^{18}}-\frac{576\left(61697M^2+34629g_{1}q^2\right)Mg_{1}q^6}{5r^{19}}\right.\nonumber\\
&&\left. +\frac{12672\left(2797M^2+362g_{1}q^2\right)g_{1}^2q^8}{5r^{20}}-\frac{15346176Mg_{1}^3q^{10}}{5r^{21}}+\frac{2439792g_{1}^4q^{12}}{r^{22}}\right]\nonumber\\
&& +g_{2}^2\left[\frac{351q^8}{50r^{17}}-\frac{2922912Mq^8}{25r^{21}}+\frac{108\left(124759M^2+21120g_{1}q^2\right)q^8}{25r^{22}}-\frac{72\left(211865M^2+249033g_{1}q^2\right)Mq^8}{25r^{23}}\right.\nonumber\\
&&\left. +\frac{144\left(198239M^2+20066g_{1}q^2\right)g_{1}q^{10}}{25r^{24}}-\frac{17466408Mg_{1}^2q^{12}}{25r^{25}}+\frac{3505176g_{1}^3q^{14}}{25r^{26}}\right]\nonumber\\
&& -g_{2}^3\left[\frac{684288q^{12}}{125r^{26}}-\frac{4307796Mq^{12}}{125r^{27}}+\frac{108\left(56375M^2+23968g_{1}q^2\right)q^{12}}{125r^{28}}-\frac{7096788Mg_{1}q^{14}}{125r^{29}}+\frac{2046582g_{1}^2q^{16}}{125r^{30}}\right]\nonumber\\
&& +g_{2}^4\left(\frac{357048q^{16}}{625r^{32}}-\frac{364851Mq^{16}}{250r^{33}}+\frac{2041551g_{1}q^{18}}{2500r^{34}}\right)-\frac{724707g_{2}^5q^{20}}{50000r^{38}}
\end{eqnarray}

We note that the first term of the $\mathcal{O}(\beta^{2})$ correction appears up to order $\mathcal{O}(r^{-11})$ and would only correct the mass term, the linear contribution of the electric charge is present at order $\mathcal{O}(r^{-12})$ and the nonlinear terms of the electric charge appear up to order $\mathcal{O}(r^{-16})$. The difference between the linear and nonlinear electrodynamics term is still $\mathcal{O}(r^{-4})$ as observed in expressions ((\ref{SolWAco}) and ((\ref{SolWAco1}). 

\subsection{Near Horizon}\label{NeaHor}

For the near horizon solution we also proposed a solution in a power series of $\beta$ for the coefficients $a_{n}$, which allowed us to determine the branch with the correct GR limit of the coefficient $a_{1}$, of the form 

\begin{equation*}
    a_{n}=\sum_{j=0}^{j_{max}}h_{n,j}\beta^{j},
\end{equation*}

\noindent the coefficient $h_{1,j}$ with $j_{max}=2$ corresponding to the coefficient $a_{1}$ and the first coefficient of $a_{2}$ which is $h_{2,0}$ are presented in section \ref{NHSec}, by completes we present the coefficients $h_{2,j}$ with $j=1$ and $j=2$,

\begin{eqnarray}
h_{2,1}&&=-\frac{3\left(g_{2}q^4-4g_{1}q^2 {r_{+}}^4+4 {r_{+}}^6\right)}{32{r_{+}}^{24}}\left(83g_{2}^2 q^8-408g_{1}g_{2}q^6 {r_{+}}^4+680g_{2}q^4 {r_{+}}^6+432 g_{1}^2 q^4 {r_{+}}^8-1120g_{1}q^2 {r_{+}}^{10}\right.\nonumber\\
&&\left.+432 {r_{+}}^{12}\right)\\
h_{2,2}&&=-\frac{9 \left(g_{2}q^4-4g_{1}q^2 {r_{+}}^4+4 {r_{+}}^6\right)}{256 {r_{+}}^{40}}\left(5549g_{2}^4 q^{16}-64640
g_{1}g_{2}^3 q^{14} {r_{+}}^4+107932 g_{2}^3 q^{12} {r_{+}}^6+269472 g_{1}^2 g_{2}^2 q^{12} {r_{+}}^8\right.\nonumber\\
&& \left. -872656g_{1}g_{2}^2 q^{10} {r_{+}}^{10}+637808 g_{2}^2 q^8 {r_{+}}^{12}-467456 g_{1}^3g_{2}q^{10} {r_{+}}^{12}+2160448 g_{1}^2g_{2}q^8 {r_{+}}^{14}-2997120g_{1}g_{2}q^6 {r_{+}}^{16}\right.\nonumber\\
&& \left. +274688 g_{1}^4 q^8 {r_{+}}^{16}+1241664g_{2}q^4 {r_{+}}^{18}-1545984g_{1}^3 q^6 {r_{+}}^{18}+2848512g_{1}^2 q^4 {r_{+}}^{20}-1993984g_{1}q^2 {r_{+}}^{22}+433152 {r_{+}}^{24}\right)\nonumber\\
\end{eqnarray}



\begin{thebibliography}{}

\bibitem{Pich}
A. Pich, \textit{Effective Field Theory}, Lectures at the 1997 Les Houches Summer School ``Probing the Standard Model of Particle Interactions''. arXiv:hep-ph/9806303v1.

\bibitem{HeisenbergEuler}
W. Heisenberg and H. Euler, \textit{Consequences of Dirac Theory of the Positron}, Z.Phys. {\bf 98}, 714 (1936).

\bibitem{Manohar}
A. V. Manohar, \textit{Effective field theories}, Lect. Notes Phys. {\bf 479}, 311 (1997).

\bibitem{Euler}
E. Euler,\textit{Über die Streuung von Licht an Licht nach der Diracschen Theorie}, Ann. Phys., Lpz. {\bf 26}, 398 (1936).

\bibitem{Adams}
A. Adams et. al., \textit{Causality, analyticity and an IR obstruction to UV completion}, J. High Energ. Phys. {\bf 10}, 014 (2006).

\bibitem{KarplusNeuman}
R. Karplus and M. Neuman, \textit{The scattering of light by light}, Phys. Rev. {\bf 83}, (1951) 776.

\bibitem{Ostro}
M. Ostrogradsky, \textit{Mémoires sur les équations différentielles, relatives au problème des isopérimètres},     Mem.Acad.St.Petersbourg {\bf 6}, 385 (1850).

\bibitem{Lanczos}
C. Lanczos, \textit{A Remarkable Property of the Riemann-Christoffel Tensor in Four Dimensions}, Ann. Math. {\bf 39}, 842 (1938).

\bibitem{Lovelock}
D. Lovelock, \textit{The Einstein Tensor and Its Generalizations}, J. Math. Phys. {\bf 12}, 498 (1971).

\bibitem{Metsaev}
R. R. Metsaev and A. A. Tseytlin, \textit{Order $\alpha'$(two-loop) equivalence of the string equations of motion and the $\sigma$-model Weyl invariance conditions: Dependence on the dilaton and the antisymmetric tensor}, Nucl. Phys. B {\bf 293}, 385 (1987).

\bibitem{Gasperini}
M. Gasperini, M. Maggiore, and G. Veneziano \textit{Towards a non-singular pre-big-bang cosmology}, Nucl. Phys. B {\bf 494}, 315 (1997).

\bibitem{Antoniou}
G. Antoniou, A. Bakopoulos, and P. Kanti, \textit{Evasion of No-Hair Theorems and Novel Black-Hole Solutions in Gauss-Bonnet Theories}, Phys. Rev. Lett. {\bf 120}, 131102 (2018).

\bibitem{Aoki}
K. Aoki, and S. Tsujikawa, \textit{Coupled vector Gauss-Bonnet theories and hairy black holes}, Phys. Lett. B {\bf 843}, 138022 (2023).

\bibitem{Kubiznak}
R. A. Hennigar, D. Kubiznak, and R. B. Mann, \textit{Generalized quasitopological gravity}, Phys. Rev. D {\bf 95}, 104042 (2017).

\bibitem{BuenoCano1}
P. Bueno and P. A. Cano, \textit{Einsteinian Cubic Gravity}, Phys. Rev. D {\bf 94}, 104005 (2016).

\bibitem{DeFelice}
A. De Felice and S. Tsujikawa, \textit{Excluding static and spherically symmetric black holes in Einsteinian cubic gravity with unsuppressed higher-order curvature terms}, Phys. Lett. B {\bf 843}, 138047 (2023).

\bibitem{BuenoEFT}
P. Bueno, P. A. Cano, and R. A. Hennigar, \textit{On the stability of Einsteinian Cubic Gravity black holes in EFT}, arXiv:2306.02924.

\bibitem{JimenezCano}
J. Beltrán Jiménez, and A. Jiménez-Cano, \textit{On the physical viability of black hole solutions in Einsteinian Cubic Gravity and its generalisations}, Phys. Dark Univ. {\bf 43}, 101387 (2024).

\bibitem{Hawking}
S. W. Hawking and T. Hertog, \textit{Living with ghosts}, Phys. Rev. D {\bf 65}, 103515 (2002).

\bibitem{Battista}
E. Battista, \textit{Quantum Schwarzschild geometry in effective field theory models of gravity}, Phys. Rev. D {\bf 109}, 026004 (2024).

\bibitem{Akiyama1}
K. Akiyama, et. al. (The Event Horizon Telescope Collaboration), \textit{First M87 Event Horizon Telescope Results. I. The Shadow of the Supermassive Black Hole}, ApJL {\bf 875}, L1 (2019).

\bibitem{Akiyama2}
K. Akiyama, et. al. (The Event Horizon Telescope Collaboration), \textit{First M87 Event Horizon Telescope Results. IV. Imaging the Central Supermassive Black Hole}, ApJL {\bf 875}, L4 (2019).

\bibitem{Akiyama3}
K. Akiyama, et. al. (The Event Horizon Telescope Collaboration), \textit{First Sagittarius A* Event Horizon Telescope Results. I. The Shadow of the Supermassive Black Hole in the Center of the Milky Way}, ApJL {\bf 930}, L12 (2022).

\bibitem{Akiyama4}
K. Akiyama, et. al. (The Event Horizon Telescope Collaboration), \textit{First Sagittarius A* Event Horizon Telescope Results. III. Imaging of the Galactic Center Supermassive Black Hole}, ApJL {\bf 930}, L14 (2022).

\bibitem{LIGO}
B. P. Abbott et al. (LIGO Scientific Collaboration and Virgo Collaboration), \textit{Observation of Gravitational Waves from a Binary Black Hole Merger}, Phys. Rev. Lett. {\bf 116}, 061102 (2016).

\bibitem{Plebanski}
J. Pleba\'nski, \textit{Lectures on Non-linear Electrodynamics}, (NORDITA, Copenhagen, 1970).

\bibitem{SalazarGarciaPlebanski}
H. Salazar, A. Garc\'ia and J. F. Pleba\'nski, \textit{Duality rotations and type D solutions to Einstein
equations with nonlinear electrodynamics sources}, J. Math. Phys. {\bf 28}, 2171 (1987).

\bibitem{MagosBreton}
D. Magos and N. Breton, \textit{Thermodynamics of the Euler-Heisenberg-AdS black hole}, Phys. Rev. D {\bf 102}, 084011 (2020).

\bibitem{AmaroMacias}
D. Amaro and A. Mac\'ias, \textit{Geodesic structure of the Euler-Heisenberg static black hole}, Phys. Rev. D {\bf 102}, 104054 (2020). 

\bibitem{Yajima}
H. Yajima and T. Tamaki, \textit{Black hole solutions in Euler-Heisenberg theory}, Phys. Rev. D {\bf 63}, 064007 (2001).

\bibitem{GibbonsRasheed}
G. W. Gibbons and D. A. Rasheed, \textit{Electric-magnetic duality rotations in nonlinear electrodynamics}, Nucl. Phys. B {\bf 454}, 185 (1995).

\bibitem{NoraRot}
N. Bretón, C. Lämmerzahl, and A. Macías, \textit{Rotating black holes in the Einstein–Euler–Heisenberg theory}, Class. Quant. Grav. {\bf 36}, 235022 (2019).

\bibitem{Allahyari}
A. Allahyari, M. Khodadi, S. Vagnozzi and D. F. Mota, \textit{Magnetically charged black holes from non-linear electrodynamics and the Event Horizon Telescope}, J. High Energ. Phys. {\bf 02}, 003 (2020).

\bibitem{Stefanov}
I. Z. Stefanov, S. S. Yazadjiev and M. D. Todorov, \textit{Scalar-tensor black holes coupled to Euler-Heisenberg nonlinear electrodynamics}, Mod. Phys. Lett. A {\bf 22}, 1217 (2007).

\bibitem{GuerreroRubiera}
M. Guerrero and D. Rubiera-Garcia, \textit{Nonsingular black holes in nonlinear gravity coupled to Euler-Heisenberg electrodynamics}, Phys. Rev. D {\bf 102}, 024005 (2020).

\bibitem{BuenoCano2}
P. Bueno and P. A. Cano, \textit{Four-dimensional black holes in Einsteinian cubic gravity}, Phys. Rev. D {\bf 94}, 124051 (2016).

\bibitem{HennigarMann}
R. A. Hennigar and R. B. Mann, \textit{Black holes in Einsteinian cubic gravity}, Phys. Rev. D {\bf 95}, 064055 (2017).

\bibitem{Sajadi}
S. N. Sajadi and S. H. Hendi, \textit{Analytically approximation solution to Einstein-Cubic gravity}, Eur. Phys. J. C {\bf 82}, 675 (2022). 

\bibitem{AdairBuenoCano}
C. Adair, P. Bueno, P. A. Cano, R. A. Hennigar and R. B. Mann, \textit{Slowly rotating black holes in Einsteinian cubic gravity}, Phys. Rev. D {\bf 102}, 084001 (2020).

\bibitem{CanoPeriniguez}
P. A. Cano and D. Pere\~niguez, \textit{Extremal Rotating Black Holes in Einsteinian Cubic Gravity}, Phys. Rev. D {\bf 101}, 044016 (2020).

\bibitem{Arciniega1}
G. Arciniega, J. D. Edelstein, and L. G. Jaime, \textit{Towards geometric inflation: the cubic case}, Phys. Lett. B {\bf 802}, 135272 (2020).

\bibitem{Arciniega2}
G. Arciniega, et. al., \textit{Geometric Inflation}, Phys. Lett. B {\bf 802}, 135242 (2020).

\bibitem{Arciniega3}
G. Arciniega, et. al. \textit{Cosmic inflation without inflaton}, Int. J. Mod. Phys. D {\bf 28}, 1944008 (2019). 

\bibitem{BuenoCanoRuiperez}
P. Bueno, P. A. Cano, and A. Ruiperez, \textit{Holographic studies of Einsteinian cubic gravity}, J. High Energ. Phys. {\bf 03}, 150 (2018).

\bibitem{EdelsteinGrandi}
J.D. Edelstein, N. Grandi, and A. Rivadulla Sanchez, \textit{Holographic superconductivity in Einsteinian Cubic Gravity}, J. High Energ. Phys. {\bf 05}, 188 (2022).

\bibitem{Poshteh1}
R. A. Hennigar, M. B. J. Poshteh, and R. B. Mann, \textit{Shadows, signals, and stability in Einsteinian cubic gravity}, Phys. Rev. D {\bf 97}, 064041 (2018).

\bibitem{Poshteh2}
M. B. J. Poshteh and R. B. Mann, \textit{Gravitational lensing by black holes in Einsteinian cubic gravity}, Phys. Rev. D {\bf 99}, 024035 (2019).

\bibitem{NUTs}
P. Bueno, P. A. Cano, R. A. Hennigar and R. B. Mann, \textit{NUTs and bolts beyond Lovelock}, J. High Energ. Phys. {\bf 10}, 095 (2018).

\bibitem{Mehdizadeh}
M. R. Mehdizadeh and A. H. Ziaie, \textit{Traversable wormholes in Einsteinian cubic gravity}, Mod. Phys. Lett. A {\bf 35}, 2050017 (2019).

\bibitem{GarciaGut}
A. A. Garcia-Diaz, and G. Gutierrez-Cano, \textit{Linear superposition of regular black hole solutions of Einstein nonlinear electrodynamics}, Phys. Rev. D {\bf 100}, 064068 (2019).

\bibitem{Kocher}
P. Kocherlakota, et al. (EHT Collaboration), \textit{Constraints on black-hole charges with the 2017 EHT observations of M87$ ^{*}$}, Phys. Rev. D {\bf 103}, 104047 (2021). 

\bibitem{Vagnozzi}
S. Vagnozzi, et al., \textit{Horizon-scale tests of gravity theories and fundamental physics from the Event Horizon Telescope image of Sagittarius A$^{*}$}, Class. Quantum Grav. {\bf 40}, 165007 (2023).

\bibitem{Baker}
T. Baker, D. Psaltis, and C. Skordis, \textit{Linking Tests of Gravity On All Scales: from the Strong-Field Regime to Cosmology}, Astrophys. J. {\bf 802}, 63 (2015).

\bibitem{Perlick}
V. Perlick, and O. Y. Tsupko, \textit{Calculating black hole shadows: Review of analytical studies}, Phys. Rep. {\bf 947}, 1 (2022).

\bibitem{Virbhadra1}
K.S. Virbhadra, \textit{Distortions of images of Schwarzschild lensing}, Phys.Rev.D {\bf 106}, 064038 (2022).

\bibitem{Virbhadra2}
K.S. Virbhadra, \textit{Compactness of supermassive dark objects at galactic centers}, arXiv:2204.01792.

\bibitem{GibbonsWL}
G. W. Gibbons, and M C Werner, \textit{Applications of the Gauss-Bonnet theorem to gravitational lensing}, Class. Quant. Grav. {\bf 25}, 235009 (2008).

\bibitem{OkyayWL}
M. Okyay, and A. Övgün, \textit{Nonlinear electrodynamics effects on the black hole shadow, deflection angle, quasinormal modes and greybody factors}, JCAP {\bf 2022}, 009 (2022).

\bibitem{QMFu}
Q.-M. Fu, L. Zhao, and Y.-X, Liu, \textit{Weak deflection angle by electrically and magnetically charged black holes from nonlinear electrodynamics}, Phys. Rev. D {\bf 104}, 024033 (2021).

\end{thebibliography}
\end{document}